\documentclass[letterpaper]{article} 
\usepackage{paperstyle}  
\nocopyright 
\usepackage[hyphens]{url}  
\usepackage{graphicx} 
\usepackage{natbib}  
\usepackage{caption} 
\definecolor{dcgreen}{HTML}{006400}

\usepackage{xspace}
\newcommand{\ours}{CRAW\xspace}

\usepackage{booktabs}
\usepackage{multirow}

\usepackage{amsmath}
\usepackage{amssymb}

\usepackage{tikz}
\usetikzlibrary{arrows.meta, positioning, fit, backgrounds, calc, shadows.blur}

\title{CRAW: Codec Robust Audio Watermarking}

\author{
    David Chernin\textsuperscript{\rm 1},
    Ethan Fetaya\textsuperscript{\rm 1,\rm 2}
}
\affiliations{
    \textsuperscript{\rm 1}Bar-Ilan University, Ramat Gan, Israel\\
    \textsuperscript{\rm 2}NVIDIA\\
    chernid@biu.ac.il, ethan.fetaya@biu.ac.il
}

\usepackage{placeins}


\begin{document}

\maketitle

\begin{abstract}
Recent advances in generative speech models have made it increasingly difficult to distinguish authentic from synthetic audio, enabling new forms of fraud and misinformation. Audio watermarking offers a promising defense by embedding an imperceptible signal into generated speech that can later be detected to verify its provenance. However, recent studies have shown that existing post-hoc watermarking methods fail under neural codecs and denoisers, transformations routinely applied during real-world storage, transmission, and processing, severely limiting their practical utility. Here we introduce \ours, a codec-robust audio watermarking framework that jointly improves robustness against neural re-synthesis while maintaining high perceptual quality. \ours combines distortion-aware training with an attention-based pooling mechanism, inference-time perceptual masking, and an error-correcting code to recover the fidelity lost during robust training. Experiments demonstrate that \ours achieves state-of-the-art robustness against neural codecs, denoisers, and vocoders while maintaining perceptual quality comparable to existing post-hoc watermarking methods. The code is available at \url{https://github.com/DavidC1212/craw}.

\end{abstract}


\section{Introduction}
\label{sec:introduction}
Recent advancements in voice cloning, the ability to generate speech that impersonates a specific speaker, have enabled increasingly convincing synthetic speech, but have also been misused for fraud and the spread of misinformation (Shaaban et al.~\citeyear{Shaaban2023audiodeepfake}). As these systems become more capable, reliably distinguishing authentic recordings from AI-generated speech has become an important challenge with significant societal implications. One promising approach is audio watermarking, which embeds an imperceptible signal into generated speech that can later be detected to verify its provenance. This can be done during generation or post-hoc on a given audio sample.\\

For watermarking to be practical, the embedded signal must remain detectable after the audio undergoes both malicious modifications and benign processing. In practice, audio is rarely stored or transmitted in its original form: voice calls, video conferencing, and online platforms routinely process speech using codecs, denoisers, and other enhancement models before delivery. Recent work by~\citet{OReilly2025shallow} demonstrated that although existing post-hoc watermarking methods remain robust to many traditional distortions, such as waveform dropout and time stretching, they fail almost completely under neural codecs and denoisers. This exposes a substantial gap between benchmark performance and real-world deployment.\\

In this work, we address this limitation with \ours (\textbf{C}odec \textbf{R}obust \textbf{A}udio \textbf{W}atermarking), a post-hoc watermarking framework designed specifically to withstand neural re-synthesis attacks while preserving perceptual audio quality. Building on TimbreWatermark~\citep{Liu2023timbre}, we first strengthen the training distortion layer, which distorts the audio signal before it reaches the detector during training, by incorporating neural codec re-synthesis and other realistic attacks, substantially improving robustness to these transformations. However, training against this broader attack distribution introduces a significant degradation in perceptual fidelity.\\

\ours addresses this robustness-fidelity trade-off through four complementary components. First, we broaden the training distortion layer to include neural codec resynthesis, closing the codec-robustness gap at a cost to fidelity. Second, we replace TimbreWatermark's uniform temporal averaging with an attention-based pooling mechanism inspired by Q-Former~\citep{Li2023blip2}, allowing the embedding and detection to be more targeted. Third, we introduce an inference-time masking strategy that selectively removes the watermark from spectrogram regions that contribute most to perceptual degradation, measured by PESQ, recovering much of the lost audio quality. Finally, we incorporate an error-correcting code to compensate for the robustness reduction introduced by masking, producing a watermarking system that simultaneously achieves high robustness and high perceptual fidelity. Concretely, our contributions are:

\begin{itemize}
\item We introduce \ours, a post-hoc audio watermarking framework designed for robustness against neural codecs, denoisers, vocoders, and other neural re-synthesis attacks that significantly degrade existing watermarking methods.
\item We propose an attention-based pooling mechanism, inspired by Q-Former, that substantially improves perceptual fidelity while maintaining robustness.
\item We introduce a fidelity-aware inference strategy combining perceptual-gradient masking with error-correcting codes to recover audio quality without sacrificing robustness.
\item We demonstrate through extensive experiments that \ours achieves state-of-the-art robustness across a wide range of neural and traditional distortions while maintaining perceptual quality comparable to existing watermarking methods.
\end{itemize}

\section{Related Work}
\label{sec:related_work}

In the post-hoc paradigm, a watermark is embedded into audio
after it's created. TimbreWatermark~\citep{Liu2023timbre}
repeats a short-time-Fourier-transform message along the time axis
and trains against a distortion layer simulating voice-conversion and
TTS pipelines; AudioSeal~\citep{SanRoman2024audioseal} adds a learned
waveform signature with a jointly trained sample-level detector for
per-timestep localization; WavMark~\citep{Chen2023wavmark} uses an
invertible network with a brute-force positional search in place of a
synchronization code. More recent methods attack the same weakness
differently: AWARE~\citep{Pavlovic2025aware} replaces
attack-simulation training with adversarial optimization under a
perceptual budget and a desynchronization-robust detector;
WMCodec~\citep{Zhou2025wmcodec} trains the
watermark jointly with a neural codec's compression and
reconstruction rather than as a separate stage. Recently, \citet{OReilly2025shallow} and
RAW-Bench~\citep{Ozer2025rawbench} both found that post-hoc methods collapse
under low-bitrate neural codecs regardless of embedding scheme, with
\citet{OReilly2025shallow} additionally showing the same collapse
under denoisers.\\

In-generation methods embed the watermark inside the
synthesis process itself, trading generality for robustness by
construction: they cannot watermark audio that already exists, and
require retraining or fine-tuning the generator. GROOT~\citep{Liu2024groot}
conditions a diffusion-based vocoder's reverse process on the
watermark as a latent variable; SOLIDO~\citep{Li2025solido} fine-tunes
a speech diffusion model via low-rank adaptation with a decoder built
for variable-length audio; Smark~\citep{Zhang2025smark} embeds into
the low-frequency wavelet sub-bands of the reverse diffusion process,
generalizing across diffusion architectures.
While \citet{OReilly2025shallow} only tested post-hoc watermarking methods, they conjecture that in-generation methods would probably have the same robustness limitation due to the low-magnitude of the watermarked signal.


\section{Background}
\label{sec:background}

\ours builds on the TimbreWatermark framework~\citep{Liu2023timbre}, whose embedder,
extractor, and training objective we summarize here. Section~\ref{sec:method}
describes the four components \ours adds on top of this architecture to address its limitations.

\paragraph{\textbf{Embedding}}
Given the carrier waveform $\mathbf{x}\in\mathbb{R}^L$, TimbreWatermark
applies the Short-Time Fourier Transform to obtain a spectrogram and
phase, each of shape $F\times T$ for $F$ frequency bins and $T$
(flexible) time frames,
\begin{equation}
S, P = \mathrm{STFT}(\mathbf{x}), \qquad S, P \in \mathbb{R}^{F\times T}.
\end{equation}
The magnitude spectrogram $S$ is the carrier: the embedder operates on
this spectrogram, not the waveform itself, and $S$ is passed through the
carrier encoder to obtain carrier features,
\begin{equation}
f_c = \mathrm{EN_c}(S) \in \mathbb{R}^{C_v\times F\times T}.
\end{equation}
The message $\mathbf{m}\in\{0,1\}^K$ is
passed through a watermark encoder,
\begin{equation}
f_m = \mathrm{EN_w}(\mathbf{m}) \in \mathbb{R}^{C_w\times F\times 1}.
\label{eq:bg_wm_encoder}
\end{equation}
Since $S$ has a flexible temporal length, $f_m$ is repeated along the
time axis; a skip connection also carries $S$ itself forward so the
embedder keeps direct access to the original carrier, and all three are
concatenated along the channel axis to form the embedder's input,
\begin{equation}
f_+ = \mathrm{Concat}\big(f_c,\ S,\ \mathrm{Repeat}(f_m, T)\big)
\in \mathbb{R}^{(C_v+1+C_w)\times F\times T}.
\label{eq:bg_concat}
\end{equation}
The watermark embedder produces the watermarked spectrogram,
\begin{equation}
S_w = \mathrm{EM}(f_+) \in \mathbb{R}^{F\times T},
\end{equation}
and the inverse STFT reconstructs the watermarked waveform using the
original phase,
\begin{equation}
\mathbf{x}_w = \mathrm{ISTFT}(S_w, P) \in \mathbb{R}^L.
\end{equation}

\paragraph{\textbf{Extraction}}
The received waveform $\tilde{\mathbf{x}}\in\mathbb{R}^L$ is converted
back to the spectrogram domain,
\begin{equation}
\tilde{S}, \tilde{P} = \mathrm{STFT}(\tilde{\mathbf{x}}), \qquad
\tilde{S},\tilde{P}\in\mathbb{R}^{F\times T}.
\end{equation}
As in the embedding phase, only the magnitude spectrogram is used. The watermark
extractor produces per-frame features,
\begin{equation}
f_w' = \mathrm{EX}(\tilde{S}) \in \mathbb{R}^{F\times T}.
\end{equation}
These per-frame features are pooled into a single vector by simple
averaging across time,
\begin{equation}
\bar{f}_w = \mathrm{Average}(f_w') \in \mathbb{R}^{F},
\label{eq:bg_pooling}
\end{equation}
 and the pooled feature is passed through the message
decoder to recover the message,
\begin{equation}
\hat{\mathbf{m}} = \mathrm{DE}(\bar{f}_w) \in \{0,1\}^K.
\end{equation}

\paragraph{\textbf{Distortion Layer}}
To make the watermarking robust to how watermarked audio might be altered
before it is verified, TimbreWatermark inserts a distortion layer
$\mathrm{DP}(\cdot)$ between embedding and extraction, targeting
voice-cloning-style resynthesis: it renormalizes the watermarked
waveform by its peak amplitude, converts it to a mel-spectrogram via
$\mathrm{Mel}(\cdot)$, and reconstructs a waveform from that
mel-spectrogram with the Griffin-Lim algorithm $\mathrm{GL}(\cdot)$,
mimicking the mel-to-waveform step of a typical voice-cloning pipeline,
\begin{equation}
\tilde{\mathbf{x}} = \mathrm{DP}(\mathbf{x}_w) =
\mathrm{GL}\!\left(\mathrm{Mel}\!\left(\frac{\mathbf{x}_w}{\max(|\mathbf{x}_w|)}\right)\right)
\in \mathbb{R}^L.
\label{eq:bg_distortion}
\end{equation}
The same extraction pipeline above runs on both the clean watermarked
audio $\mathbf{x}_w$ and its distorted counterpart $\tilde{\mathbf{x}}$
each step, so the detector learns to recover the message from either.

\paragraph{\textbf{Losses}}
Fidelity is enforced with an MSE loss between the clean and watermarked
waveforms,
\begin{equation}
\mathcal{L}_e = \frac{1}{L}\sum_{i=1}^{L} \big((\mathbf{x}_w)_i - \mathbf{x}_i\big)^2,
\label{eq:bg_fidelity_loss}
\end{equation}
plus an adversarial loss against a
discriminator $D$ trained to distinguish $\mathbf{x}_w$ from clean audio,
\begin{equation}
\mathcal{L}_{\text{adv}} = -\log\big(\sigma(D(\mathbf{x}_w))\big),
\label{eq:bg_adv_loss}
\end{equation}
where $\sigma(\cdot)$ is the sigmoid function. $D$ itself is trained
each step with its own loss via a separate optimizer,
\begin{equation}
\mathcal{L}_d = -\log(\sigma(D(\mathbf{x}))) - \log(1-\sigma(D(\mathbf{x}_w))).
\label{eq:bg_disc_loss}
\end{equation}
Since extraction runs on both the clean and distorted branches
(Eq.~\ref{eq:bg_distortion}), we also get a message loss on each,
\begin{equation}
\mathcal{L}_{\text{msg}} = \frac{1}{K}\sum_{i=1}^{K} (\hat{m}_i - m_i)^2,
\label{eq:bg_msg_loss}
\end{equation}
with $\mathcal{L}_{\text{msg}}$ computed on the distorted branch and
$\hat{\mathcal{L}}_{\text{msg}}$ (of the same form) on the clean branch.
The total training objective is given by
\begin{equation}
\mathcal{L} = \lambda_e \mathcal{L}_e + \lambda_a \mathcal{L}_{\text{adv}}
+ \lambda_m\big(\mathcal{L}_{\text{msg}} + \hat{\mathcal{L}}_{\text{msg}}\big).
\label{eq:bg_total_loss}
\end{equation}

\section{Method}
\label{sec:method}
\ours adds four components to fix robustness and maintain fidelity. The entire pipeline is shown in Figure~\ref{fig:training}.
The original distortion layer only simulates voice-cloning-style
vocoder resynthesis, so we first broaden it to cover additional distortions, including codecs
(Section~\ref{subsec:distortion}). This closes the codec-robustness
gap, but at a significant cost to fidelity. Q-Former pooling (Section~\ref{subsec:qformer})
and masking (Section~\ref{subsec:inference_masking}) then regain that fidelity:
Q-Former allows the detector to down-weight corrupted frames instead
of averaging them in uniformly, and masking withholds the watermark
from the spectrogram cells that cost the most, together recovering most
of the fidelity loss. An error-correcting code
(Section~\ref{subsec:ecc_addition}) is added last, recovering the codec
robustness Q-Former and masking alone give up, at a negligible further
cost to fidelity. The embedder, detector, Q-Former, and ECC are trained
jointly (Eq.~\ref{eq:bg_total_loss}); masking is applied only
afterward, at inference, and does not enter this objective.

\definecolor{craftBlue}{HTML}{6366F1}
\definecolor{craftBlueFill}{HTML}{EEF2FF}
\definecolor{craftAmber}{HTML}{D97706}
\definecolor{craftAmberFill}{HTML}{FFFBEB}
\definecolor{craftSlate}{HTML}{475569}
\definecolor{craftSlateFill}{HTML}{F1F5F9}
\definecolor{craftRose}{HTML}{DC2626}
\definecolor{craftRoseFill}{HTML}{FEF2F2}
\definecolor{craftGold}{HTML}{CA8A04}
\definecolor{craftGoldFill}{HTML}{FEF9C3}
\definecolor{craftTeal}{HTML}{0D9488}
\definecolor{craftTealFill}{HTML}{F0FDFA}
\definecolor{craftMist}{HTML}{818CF8}
\definecolor{craftMistFill}{HTML}{F5F6FE}
\definecolor{craftOrange}{HTML}{C2410C}
\definecolor{craftOrangeFill}{HTML}{FFEDD5}
\definecolor{craftPanelBlue}{HTML}{F7F8FF}
\definecolor{craftPanelRose}{HTML}{FFFAFA}
\definecolor{craftPanelSlate}{HTML}{FAFBFC}
\definecolor{craftGreen}{HTML}{059669}
\definecolor{craftGreenFill}{HTML}{ECFDF5}
\definecolor{craftPanelGreen}{HTML}{F6FEFA}
\definecolor{craftCyan}{HTML}{0891B2}
\definecolor{craftCyanFill}{HTML}{ECFEFF}
\definecolor{craftViolet}{HTML}{7C3AED}
\definecolor{craftVioletFill}{HTML}{F5F3FF}
\definecolor{craftPanelTeal}{HTML}{F7FEFD}

\begin{figure*}[t]
\centering
\resizebox{\textwidth}{!}{%
\begin{tikzpicture}[
  font=\sffamily\small,
  >=Stealth,
  block/.style={draw=craftBlue,semithick,rounded corners=4pt,fill=craftBlueFill,
    minimum width=26mm,minimum height=14mm,align=center,inner sep=2.4mm,
    font=\sffamily\large\bfseries,
    blur shadow={shadow blur steps=6,shadow xshift=0.5pt,shadow yshift=-0.7pt,
      shadow blur radius=2.2pt}},
  sblock/.style={draw=craftAmber,semithick,rounded corners=4pt,fill=craftAmberFill,
    minimum width=26mm,minimum height=14mm,align=center,inner sep=2.4mm,
    font=\sffamily\large\bfseries,
    blur shadow={shadow blur steps=6,shadow xshift=0.5pt,shadow yshift=-0.7pt,
      shadow blur radius=2.2pt}},
  lblock/.style={draw=craftGreen,semithick,rounded corners=4pt,fill=craftGreenFill,
    minimum width=26mm,minimum height=14mm,align=center,inner sep=2.4mm,
    font=\sffamily\large\bfseries,
    blur shadow={shadow blur steps=6,shadow xshift=0.5pt,shadow yshift=-0.7pt,
      shadow blur radius=2.2pt}},
  proc/.style={draw=craftMist,thin,rounded corners=3mm,fill=craftMistFill,
    minimum width=19mm,minimum height=8.5mm,align=center,inner sep=1.6mm,
    font=\sffamily\small,text=craftMist!55!black},
  attackproc/.style={draw=craftOrange,semithick,rounded corners=4pt,fill=craftOrangeFill,
    minimum width=18mm,minimum height=10mm,align=center,inner sep=1.4mm,
    font=\sffamily\normalsize\bfseries,text=craftOrange},
  mergept/.style={draw=craftMist,semithick,circle,minimum size=7mm,inner sep=0pt,
    fill=craftMistFill,font=\sffamily\bfseries\normalsize,text=craftMist!55!black},
  imgn/.style={align=center,inner sep=0pt},
  arr/.style={-{Stealth[length=2.6mm,width=2.2mm]},semithick,draw=gray!45!black},
  cellGold/.style={draw=craftGold,semithick,rounded corners=1.5pt,fill=craftGoldFill,
    minimum width=0.57cm,minimum height=0.57cm,inner sep=1pt,
    font=\scriptsize\bfseries\itshape},
  cellCyan/.style={draw=craftCyan,semithick,rounded corners=1.5pt,fill=craftCyanFill,
    minimum width=0.57cm,minimum height=0.57cm,inner sep=1pt,
    font=\scriptsize\bfseries\itshape},
  cellViolet/.style={draw=craftViolet,semithick,rounded corners=1.5pt,fill=craftVioletFill,
    minimum width=0.57cm,minimum height=0.57cm,inner sep=1pt,
    font=\scriptsize\bfseries\itshape},
]

\def\bitvec{%
  \begin{tikzpicture}[baseline=(bb.center),node distance=1pt]
    \node[cellGold](b0){$b_0$};
    \node[cellGold,below=2pt of b0](b1){$b_1$};
    \node[font=\scriptsize,below=1pt of b1](bd){$\vdots$};
    \node[cellGold,below=1pt of bd](bn){$b_n$};
    \coordinate(bb)at($(b0)!0.5!(bn)$);
    \draw[rounded corners=3pt,thick]
      ([xshift=-3pt,yshift=3pt]b0.north west)rectangle
      ([xshift=3pt,yshift=-3pt]bn.south east);
  \end{tikzpicture}}

\def\repmat{%
  \begin{tikzpicture}[baseline=(mm.center)]
    \foreach \xoff/\lbl in {0cm/1,0.65cm/2,1.30cm/t}{
      \node[cellViolet]at(\xoff, 0.00cm){$f_0^{\lbl}$};
      \node[cellViolet]at(\xoff,-0.65cm){$f_1^{\lbl}$};
      \node[font=\scriptsize]at(\xoff,-1.25cm){$\vdots$};
      \node[cellViolet]at(\xoff,-1.78cm){$f_l^{\lbl}$};
    }
    \node[font=\normalsize]at(1.78cm,-0.89cm){$\cdots$};
    \coordinate(mm)at(0.65cm,-0.89cm);
    \draw[rounded corners=3pt,thick]
      (-0.33cm,0.33cm)rectangle(2.00cm,-2.12cm);
  \end{tikzpicture}}

\def\codewordc{%
  \begin{tikzpicture}[baseline=(bb.center),node distance=1pt]
    \node[cellCyan](cc0){$c_0$};
    \node[cellCyan,below=2pt of cc0](cc1){$c_1$};
    \node[font=\scriptsize,below=1pt of cc1](ccd){$\vdots$};
    \node[cellCyan,below=1pt of ccd](ccn){$c_N$};
    \coordinate(bb)at($(cc0)!0.5!(ccn)$);
    \draw[rounded corners=3pt,thick]
      ([xshift=-3pt,yshift=3pt]cc0.north west)rectangle
      ([xshift=3pt,yshift=-3pt]ccn.south east);
  \end{tikzpicture}}

\def\codewordchat{%
  \begin{tikzpicture}[baseline=(bb.center),node distance=1pt]
    \node[cellCyan](ch0){$\hat{c}_0$};
    \node[cellCyan,below=2pt of ch0](ch1){$\hat{c}_1$};
    \node[font=\scriptsize,below=1pt of ch1](chd){$\vdots$};
    \node[cellCyan,below=1pt of chd](chn){$\hat{c}_N$};
    \coordinate(bb)at($(ch0)!0.5!(chn)$);
    \draw[rounded corners=3pt,thick]
      ([xshift=-3pt,yshift=3pt]ch0.north west)rectangle
      ([xshift=3pt,yshift=-3pt]chn.south east);
  \end{tikzpicture}}


\node[imgn]  at(-10.5,-4.2)(x)    {\includegraphics[width=1.8cm,height=1cm]{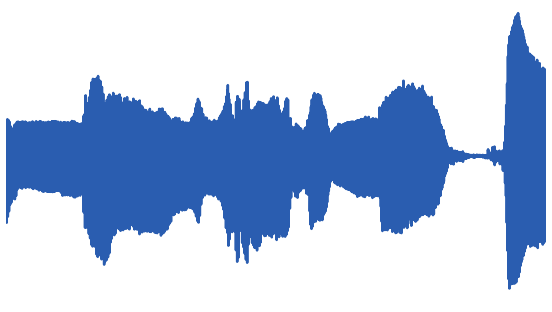}\\[1mm]\footnotesize Waveform $x$};
\node[inner sep=0pt] at(-7,   -4.2)(stftPt){};
\node[block] at(-3.5, -4.2)(ENc)  {Carrier\\Encoder};
\node[mergept]at( 0,  -4.2)(cat)  {$+$};
\node[block] at( 3.5, -4.2)(EM)   {Watermark\\Embedder};
\node[inner sep=0pt] at( 7,  -4.2)(istftPt){};
\node[imgn]  at(10.5,-4.2)(xw)
{\includegraphics[width=1.8cm,height=1cm]{waveform_icon.png}\\[1mm]\footnotesize Watermarked $x_w$};
\node[block] at(10.5,-2.2)(disc){Discriminator\\};

\node[imgn]  at(-10.5,-1.5)(m)    {\bitvec\\[1mm]\footnotesize Message $\mathbf{m}$};
\node[imgn]  at(-8,   -1.5)(codec){\codewordc\\[1mm]\footnotesize Codeword $\mathbf{c}$};
\node[sblock]at(-5,   -1.3)(wmEnc){Watermark\\Encoder};
\node[imgn]  at(0,-1.3)(rep){\repmat};

\node[attackproc]at(10.5,-9.4)(dgym) {%
  {\ttfamily\fontsize{5.5}{6.5}\selectfont [noise, codec, mp3 $\cdots$]}\\[0.6mm]
  {\normalsize Distortion Layer}};
\node[imgn]  at( 7,  -9.4)(ydist){\includegraphics[width=1.8cm,height=1cm]{waveform_icon.png}\\[1mm]\footnotesize Distorted $\tilde{x}$};
\node[sblock]at( 3.5,-9.4)(EX)   {Watermark\\Extractor};
\node[lblock]at( 0,  -9.4)(pool) {Q-Former\\Pooling};
\node[imgn] at(-2.5,-9.4)(poolvec){%
  \begin{tikzpicture}[baseline=(bb.center),node distance=1pt]
    \node[cellViolet](pv0){$f_0$};
    \node[cellViolet,below=2pt of pv0](pv1){$f_1$};
    \node[font=\scriptsize,below=1pt of pv1](pvd){$\vdots$};
    \node[cellViolet,below=1pt of pvd](pvl){$f_l$};
    \coordinate(bb)at($(pv0)!0.5!(pvl)$);
    \draw[rounded corners=3pt,thick]
      ([xshift=-3pt,yshift=3pt]pv0.north west)rectangle
      ([xshift=3pt,yshift=-3pt]pvl.south east);
  \end{tikzpicture}};
\node[sblock]at(-5,-9.4)(wmDec){Watermark\\Decoder};
\node[imgn]  at(-8,  -9.4)(codehat){\codewordchat\\[1mm]\footnotesize Codeword $\hat{\mathbf{c}}$};
\node[imgn]  at(-10.5,-9.4)(mhat){\bitvec\\[1mm]\footnotesize Decoded $\hat{\mathbf{m}}$};


\draw[arr](x)--(ENc)node[midway,above,font=\scriptsize]{STFT};
\draw[arr](ENc)--(cat);
\draw[arr](cat)--(EM);
\draw[arr](EM)--(istftPt)node[midway,above,font=\scriptsize]{ISTFT};
\draw[arr](istftPt)--(xw);
\draw[arr,gray](xw.north)--(disc.south);
\draw[gray,dashed](disc.north)--++(0,0.5)node[above,font=\scriptsize,gray](cleanq){clean or watermarked?};

\draw[arr](stftPt)--++(0,-0.85)-|(cat.south);
\draw[arr](stftPt)--(-7,-5.6)--(7,-5.6)--(istftPt);
\node[font=\sffamily\scriptsize,text=craftMist!55!black,above] at(0,-5.6){Phase};

\draw[arr](m)--(codec)node[midway,above,font=\scriptsize]{ECC Encode};
\draw[arr]($(codec)+(0.39,0.2)$)--(wmEnc.west);
\node[imgn] at(-2.5, -1.3)(fvec){%
  \begin{tikzpicture}[baseline=(bb.center),node distance=1pt]
    \node[cellViolet](f0){$f_0$};
    \node[cellViolet,below=2pt of f0](f1){$f_1$};
    \node[font=\scriptsize,below=1pt of f1](fd){$\vdots$};
    \node[cellViolet,below=1pt of fd](fl){$f_l$};
    \coordinate(bb)at($(f0)!0.5!(fl)$);
    \draw[rounded corners=3pt,thick]
      ([xshift=-3pt,yshift=3pt]f0.north west)rectangle
      ([xshift=3pt,yshift=-3pt]fl.south east);
  \end{tikzpicture}};

\draw[arr](wmEnc.east)--(fvec.west);
\draw[arr](fvec.east)--node[above,font=\scriptsize]{repeat}(rep.west);
\draw[arr](rep.south)--(cat.north);

\draw[arr](xw.south)--(dgym.north);

\draw[arr](dgym.west)--(ydist.east);
\draw[arr](ydist.west)--(EX.east)node[midway,above,font=\scriptsize]{STFT};
\draw[arr](EX.west)--(pool.east);
\draw[arr](pool.west)--(poolvec.east);
\draw[arr](poolvec.west)--(wmDec.east);
\draw[arr](wmDec.west)--(codehat.east);
\draw[arr](codehat.west)--(mhat.east)node[midway,above,font=\scriptsize]{ECC Decode};

\coordinate(topE)   at(0,-0.3);
\coordinate(botE)   at(0,-5.9);
\coordinate(topDet) at([yshift=6mm]EX.north);
\coordinate(botDet) at([yshift=-6mm]EX.south);

\begin{scope}[on background layer]
  \node[rounded corners=10pt,fill=craftPanelBlue,draw=craftBlue!25,thin,
    inner sep=7mm,fit=(x)(xw)(m)(rep)(disc)(cleanq)(topE)(botE)](embPanel){};
  \node[rounded corners=10pt,fill=craftPanelGreen,draw=craftGreen!25,thin,
    inner sep=7mm,fit=(mhat)(ydist)(topDet)(botDet)](detPanel){};
\end{scope}
\node[anchor=north west,font=\sffamily\scriptsize\bfseries,inner sep=0pt,
  text=craftBlue] at([xshift=3mm,yshift=-2.5mm]embPanel.north west)
  {Embedder};
\node[anchor=north west,font=\sffamily\scriptsize\bfseries,inner sep=0pt,
  text=craftGreen] at([xshift=3mm,yshift=-2.5mm]detPanel.north west)
  {Detector};

\end{tikzpicture}}
\caption{\ours training pipeline. The embedder encodes
message $\mathbf{m}$ via ECC and embeds it into carrier speech $\mathbf{x}$.
The distortion layer simulates attacks during training.
The detector recovers $\hat{\mathbf{m}}$ via
Q-Former pooling and ECC decoding.}
\label{fig:training}
\end{figure*}

\subsection{Distortion Layer}
\label{subsec:distortion}

The original distortion layer (Eq.~\ref{eq:bg_distortion}) is never trained
against codec perturbation, so it is not surprising that it offers no codec
robustness (Table~\ref{tab:main_robustness}). We
broaden its attack pool to include neural codec resynthesis via
FACodec~\citep{Ju2024naturalspeech3} alongside classical
signal-processing perturbations. This allows the embedder and detector to cope
with the attacks seen during training and generalize to unseen members
of the same attack family. Each step, an attack is drawn according to a
fixed per-attack probability,
\begin{equation}
\tilde{\mathbf{x}} = \mathcal{A}(\mathbf{x}_w), \qquad
\mathcal{A} \sim p(\mathcal{A}), \quad \mathcal{A}\in\{\mathcal{A}_1,\dots,\mathcal{A}_N\}.
\end{equation}
The exact perturbations used and their sampling probabilities are
detailed in the supplementary material.
\subsection{Q-Former Pooling}
\label{subsec:qformer}

Broadening the distortion layer forces the watermark embedding to withstand a harder, more diverse attack pool. However, this comes at a high cost to fidelity, which must be mitigated for this watermarking approach to be practical.  We first notice that features are pooled over frames into a single vector by simple averaging
across time (Eq.~\ref{eq:bg_pooling}), which weighs every frame equally. This pushes the embedding to appear everywhere, even when it conflicts with the reconstruction loss. We instead propose to pool
with a mechanism inspired by Q-Former~\citep{Li2023blip2}: a single
learnable query $q\in\mathbb{R}^{1\times F}$ cross-attends over the $T$
frames of $f_w' \in \mathbb{R}^{F\times T}$, treated here as $T$ tokens
$f_w'^\top \in \mathbb{R}^{T\times F}$, letting the extractor select which timesteps to extract the embedding from. The query, key, and value projections
$W_Q, W_K, W_V \in \mathbb{R}^{F\times F}$ are split into $h$ parallel
attention heads, each outputs a $d$-dimensional subspace
($d{=}F/h$) and attending independently, with head $j\in\{1,\dots,h\}$
using its own slice $W_Q^{(j)}, W_K^{(j)}, W_V^{(j)} \in
\mathbb{R}^{F\times d}$.
\begin{equation}
\alpha_j = \mathrm{softmax}\!\left(\frac{(qW_Q^{(j)})(f_w'^\top W_K^{(j)})^\top}{\sqrt{d}}\right) \in \mathbb{R}^{1\times T},
\label{eq:qformer_attn_weights}
\end{equation}
\begin{equation}
\mathrm{head}_j = \alpha_j f_w'^\top W_V^{(j)} \in \mathbb{R}^{d}.
\label{eq:qformer_head}
\end{equation}
The heads are concatenated and passed through an output projection
$W_O \in \mathbb{R}^{F\times F}$ back to dimension $F$,
\begin{equation}
\bar{f}_w = \mathrm{Concat}(\mathrm{head}_1, \dots, \mathrm{head}_h)\, W_O \in \mathbb{R}^{F},
\label{eq:qformer_attn}
\end{equation}
followed by a feed-forward block with a residual connection, as is
standard in transformer attention blocks. This pooled feature replaces
Background's averaged $\bar{f}_w$ (Eq.~\ref{eq:bg_pooling}) at the same
output dimension $F$.

\subsection{Inference-Time Masking}
\label{subsec:inference_masking}
\begin{figure*}[t]
\centering
\includegraphics[width=0.85\textwidth]{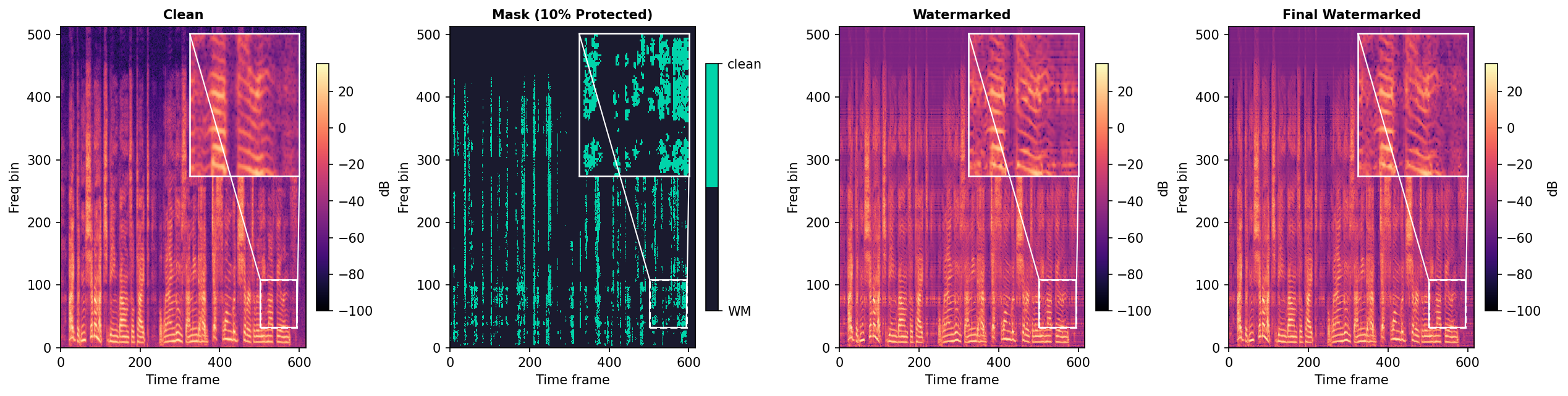}
\caption{Masking on a real clip from the test set ($k{=}10\%$). Left to
right: the clean spectrogram $S$, the binary mask $M$ (teal pixels are
kept clean; dark pixels carry the watermark), the watermarked
spectrogram $S_w$ before masking, and the final spectrogram
$S_w'$. Insets zoom into a harmonic burst: the mask protects the loud
region and pushes the watermark into the quieter gaps between bursts.}
\label{fig:mask_closeup}
\end{figure*}

The Q-Former pooling recovers much of the fidelity, but does not close the gap completely. We further propose inference-time masking, applied after the embedder has already produced $\mathbf{x}_w$, to recover the rest. To find the regions of the spectrogram most responsible for perceptual degradation, we reconstruct the clean spectrogram $S$ back to a waveform via the inverse STFT so it can be compared against $\mathbf{x}_w$ under a differentiable PESQ
implementation.\footnote{We use \texttt{torch-pesq}: \url{https://github.com/audiolabs/torch-pesq}.}
We then take the gradient of the resulting PESQ loss with respect to
$S$ itself: since we want to increase the watermarked audio's PESQ
score, this gradient marks exactly the pixels most responsible for
degrading it, giving a per-pixel sensitivity map,
\begin{equation}
G = \left|\frac{\partial\,\mathrm{PESQ}(\mathbf{x}, \mathbf{x}_w)}{\partial S}\right|.
\end{equation}
Let $M \in \{0,1\}^{F \times T}$ mark the top-$k\%$ of the pixels by $G$:
these are the pixels the listener is most sensitive to
(Figure~\ref{fig:mask_closeup}). The final spectrogram is given by removing the watermark signal using the mask $M$.
\begin{equation}
S_w' = M \odot S + (1 - M) \odot S_w.
\end{equation}
We recover the watermarked waveform with an inverse STFT. This entire process is performed at inference time; therefore, it does not affect the training process.



\subsection{Error-Correcting Code}
\label{subsec:ecc_addition}
The main challenge in this work is to achieve both robustness and fidelity simultaneously, as changes that improve one goal hurt the other. The final change we made to recover the robustness that was degraded by the masking operation is to include an error-correcting code (ECC).
An ECC adds a further, independent form of
redundancy on top of the time repetition already present in the
embedding (Eq.~\ref{eq:bg_concat}). We quantify this in the ablation study
(Section~\ref{para:ecc}).

The message $\mathbf{m}\in\{0,1\}^K$ first
goes through an ECC which encodes $\mathbf{m}$ into a longer
codeword $\mathbf{c}\in\{0,1\}^N$,
\begin{equation}
\mathbf{c} = \mathrm{ECC}_{\mathrm{enc}}(\mathbf{m}),
\end{equation}
and the codeword $\mathbf{c}$, rather than the raw message, is what we feed and train the watermark encoder on
\begin{equation}
f_m = \mathrm{EN_w}(\mathbf{c}).
\end{equation}

Because \ours operates on the codeword rather than the raw message,
Q-Former's pooled feature $\bar{f}_w$ (Eq.~\ref{eq:qformer_attn}) is
passed through the message decoder to recover the codeword,
\begin{equation}
\hat{\mathbf{c}} = \mathrm{DE}(\bar{f}_w),
\end{equation}
and the message loss is applied at the codeword level rather than the
raw-message level,
\begin{equation}
\mathcal{L}_{\text{msg}} = \frac{1}{N}\sum_{i=1}^{N} (\hat{c}_i - c_i)^2,
\label{eq:codeword_msg_loss}
\end{equation}
where $N$ is the codeword length, with $\mathcal{L}_{\text{msg}}$ and
$\hat{\mathcal{L}}_{\text{msg}}$ both taking this form. At inference, the
ECC's decoder recovers the raw message from the decoded codeword,
\begin{equation}
\hat{\mathbf{m}} = \mathrm{ECC}_{\mathrm{dec}}(\hat{\mathbf{c}}).
\end{equation}

\section{Experiments}
\label{sec:results}
In this section, we show how our method achieves strong robustness results on a variety of perturbations while maintaining high fidelity, measured by PESQ. Most important are codec perturbations, where the best competing method achieves at most 25.2\% F1 score.


\subsection{Experimental Setup}
\label{subsec:setup}

\begin{table*}[t]
\centering
\small
\begin{tabular}{l l c c c c c}
\toprule
\textbf{Category} & \textbf{Attack} & \textbf{Baseline} & \textbf{+DL} & \textbf{+QFormer} & \textbf{+Mask} & \textbf{+ECC} \\
\midrule
\multirow{11}{*}{\textbf{Sig.-Proc.}}
 & Crop Front                & \textbf{0.998$\pm$0.000} & 0.994$\pm$0.000 & 0.964$\pm$0.009 & 0.965$\pm$0.009 & 0.994$\pm$0.000 \\
 & Crop Middle               & \textbf{0.998$\pm$0.000} & 0.994$\pm$0.000 & 0.995$\pm$0.000 & 0.996$\pm$0.000 & 0.992$\pm$0.003 \\
 & Crop Back                 & \textbf{0.998$\pm$0.000} & 0.994$\pm$0.000 & 0.993$\pm$0.003 & 0.993$\pm$0.003 & 0.992$\pm$0.003 \\
 & Resample 16kHz            & \textbf{0.998$\pm$0.000} & 0.994$\pm$0.000 & 0.995$\pm$0.000 & 0.996$\pm$0.000 & 0.994$\pm$0.000 \\
 & Resample 8kHz             & \textbf{0.995$\pm$0.002} & 0.994$\pm$0.000 & 0.995$\pm$0.000 & \textbf{0.996$\pm$0.000} & 0.994$\pm$0.000 \\
 & White Noise               & \textbf{0.998$\pm$0.000} & 0.994$\pm$0.000 & 0.995$\pm$0.000 & 0.996$\pm$0.000 & 0.994$\pm$0.000 \\
 & Amp Scale                 & \textbf{0.998$\pm$0.000} & 0.994$\pm$0.000 & 0.995$\pm$0.000 & 0.996$\pm$0.000 & 0.994$\pm$0.000 \\
 & MP3 Comp (8 kbps)         & 0.861$\pm$0.020 & 0.994$\pm$0.000 & 0.995$\pm$0.000 & \textbf{0.996$\pm$0.000} & 0.994$\pm$0.000 \\
 & Recount 8bps              & \textbf{0.998$\pm$0.000} & 0.994$\pm$0.000 & 0.995$\pm$0.000 & 0.996$\pm$0.000 & 0.994$\pm$0.000 \\
 & Median Filt (35)           & 0.893$\pm$0.016 & 0.855$\pm$0.021 & \textbf{0.982$\pm$0.005} & \textbf{0.983$\pm$0.005} & \textbf{0.976$\pm$0.007} \\
 & Time Stretch               & \textbf{0.998$\pm$0.000} & 0.994$\pm$0.000 & 0.995$\pm$0.000 & 0.996$\pm$0.000 & 0.994$\pm$0.000 \\
\midrule
\multirow{2}{*}{\textbf{Filter}}
 & LowPass 2kHz              & 0.855$\pm$0.020 & 0.994$\pm$0.000 & 0.995$\pm$0.000 & \textbf{0.996$\pm$0.000} & 0.994$\pm$0.000 \\
 & HighPass 500Hz            & \textbf{0.998$\pm$0.000} & 0.979$\pm$0.006 & 0.995$\pm$0.000 & 0.996$\pm$0.000 & 0.994$\pm$0.000 \\
\midrule
\multirow{3}{*}{\textbf{Codec}}
 & FACodec                   & 0.029$\pm$0.016 & \textbf{0.994$\pm$0.000} & 0.934$\pm$0.013 & 0.937$\pm$0.012 & 0.982$\pm$0.005 \\
 & EnCodec 6kbps$^\dagger$             & 0.077$\pm$0.026 & \textbf{0.994$\pm$0.000} & \textbf{0.993$\pm$0.003} & 0.926$\pm$0.014 & 0.982$\pm$0.006 \\
 & TiCodec 1g4r$^\dagger$              & 0.029$\pm$0.016 & 0.895$\pm$0.016 & \textbf{0.948$\pm$0.011} & 0.684$\pm$0.030 & 0.838$\pm$0.021 \\
\midrule
\multirow{4}{*}{\textbf{Denoiser}}
 & FRCRN+Denoise (0 dB)$^\dagger$       & 0.067$\pm$0.024 & 0.818$\pm$0.022 & \textbf{0.942$\pm$0.012} & \textbf{0.943$\pm$0.012} & 0.913$\pm$0.015 \\
 & FRCRN+Denoise (5 dB)$^\dagger$       & 0.372$\pm$0.040 & \textbf{0.981$\pm$0.006} & \textbf{0.987$\pm$0.004} & \textbf{0.988$\pm$0.004} & \textbf{0.989$\pm$0.003} \\
 & MossFormer+Denoise (0 dB)$^\dagger$ & 0.104$\pm$0.027 & \textbf{0.963$\pm$0.009} & \textbf{0.953$\pm$0.010} & \textbf{0.954$\pm$0.010} & \textbf{0.942$\pm$0.012} \\
 & MossFormer+Denoise (5 dB)$^\dagger$ & 0.325$\pm$0.039 & 0.989$\pm$0.004 & \textbf{0.995$\pm$0.000} & \textbf{0.996$\pm$0.000} & 0.987$\pm$0.004
 \\\midrule
\multirow{2}{*}{\textbf{Vocoder}}
 & HiFi-GAN$^\dagger$                  & 0.778$\pm$0.025 & \textbf{0.994$\pm$0.000} & \textbf{0.993$\pm$0.002} & 0.990$\pm$0.003 & \textbf{0.994$\pm$0.000} \\
 & Vocos$^\dagger$                     & \textbf{0.998$\pm$0.000} & 0.994$\pm$0.000 & 0.995$\pm$0.000 & 0.996$\pm$0.000 & 0.994$\pm$0.000 \\
\midrule
\multicolumn{2}{l}{\textbf{Worst Case}} & 0.029$\pm$0.016 & 0.818$\pm$0.022 & \textbf{0.934$\pm$0.013} & 0.684$\pm$0.030 & 0.838$\pm$0.021 \\
\bottomrule
\end{tabular}
\caption{Component ablation of \ours: robustness (F1$\pm$SE) at each incremental stage, on a 200-clip subset. $\dagger$ denotes unseen attacks.}
\label{tab:ablation_robustness}
\end{table*}

\begin{table}[t]
\centering
\small
\setlength{\tabcolsep}{3pt}
\begin{tabular}{l c c c}
\toprule
\textbf{Model} & \textbf{SI-SNR (dB) $\uparrow$} & \textbf{PESQ $\uparrow$} & \textbf{STOI $\uparrow$} \\
\midrule
Baseline      & \textbf{27.89$\pm$0.17} & 4.018$\pm$0.010 & \textbf{0.993$\pm$0.000} \\
+DL           & 14.64$\pm$0.20 & 3.061$\pm$0.014 & 0.926$\pm$0.002 \\
+QFormer      & 17.85$\pm$0.20 & 3.576$\pm$0.014 & 0.961$\pm$0.001 \\
+Mask         & 19.12$\pm$0.20 & \textbf{4.085$\pm$0.008} & 0.974$\pm$0.001 \\
+ECC (\ours)  & 18.39$\pm$0.17 & 4.005$\pm$0.011 & 0.966$\pm$0.001 \\
\bottomrule
\end{tabular}
\caption{Fidelity metrics (mean$\pm$SE) for each incremental component of \ours (200-clip subset).}
\label{tab:ablation_fidelity}
\end{table}

\begin{table*}[t]
\centering
\small
\setlength{\tabcolsep}{4pt}
\begin{tabular}{l l c c c c c c}
\toprule
\textbf{Category} & \textbf{Attack} & \textbf{WavMark} & \textbf{AudioSeal} & \textbf{TimbreWM} & \textbf{WMCodec} & \textbf{AWARE} & \textbf{\ours (Ours)} \\
\midrule
\multirow{11}{*}{\textbf{Sig.-Proc.}}
 & Crop Front & 0.014$\pm$0.003 & 0.101$\pm$0.008 & \textbf{0.992$\pm$0.000} & 0.308$\pm$0.011 & 0.022$\pm$0.004 & \textbf{0.992$\pm$0.001} \\
 & Crop Middle & 0.023$\pm$0.004 & \textbf{0.995$\pm$0.000} & 0.992$\pm$0.000 & 0.163$\pm$0.009 & 0.015$\pm$0.003 & 0.991$\pm$0.001 \\
 & Crop Back & 0.048$\pm$0.006 & \textbf{0.995$\pm$0.000} & 0.992$\pm$0.000 & 0.296$\pm$0.010 & 0.073$\pm$0.007 & 0.988$\pm$0.001 \\
 & Resample 16kHz & \textbf{1.000$\pm$0.000} & 0.995$\pm$0.000 & 0.992$\pm$0.000 & 0.995$\pm$0.000 & 0.997$\pm$0.000 & 0.995$\pm$0.000 \\
 & Resample 8kHz & \textbf{1.000$\pm$0.000} & 0.995$\pm$0.000 & 0.992$\pm$0.000 & 0.738$\pm$0.008 & 0.997$\pm$0.000 & 0.995$\pm$0.000 \\
 & White Noise & 0.394$\pm$0.011 & 0.992$\pm$0.001 & 0.991$\pm$0.001 & 0.960$\pm$0.003 & \textbf{0.996$\pm$0.000} & 0.995$\pm$0.000 \\
 & Amp Scale & \textbf{1.000$\pm$0.000} & 0.995$\pm$0.000 & 0.992$\pm$0.000 & 0.995$\pm$0.000 & 0.997$\pm$0.000 & 0.995$\pm$0.000 \\
 & MP3 Comp (8 kbps) & 0.000$\pm$0.000 & 0.858$\pm$0.005 & 0.885$\pm$0.005 & 0.039$\pm$0.005 & 0.458$\pm$0.011 & \textbf{0.994$\pm$0.000} \\
 & Recount 8bps & 0.926$\pm$0.004 & 0.995$\pm$0.000 & 0.992$\pm$0.000 & 0.993$\pm$0.001 & \textbf{0.997$\pm$0.000} & 0.995$\pm$0.000 \\
 & Median Filt (35) & 0.014$\pm$0.003 & 0.096$\pm$0.007 & 0.922$\pm$0.004 & 0.127$\pm$0.008 & 0.039$\pm$0.005 & \textbf{0.975$\pm$0.002} \\
 & Time Stretch & 0.950$\pm$0.003 & 0.982$\pm$0.002 & 0.992$\pm$0.000 & 0.888$\pm$0.005 & 0.970$\pm$0.002 & \textbf{0.995$\pm$0.000} \\
\midrule
\multirow{2}{*}{\textbf{Filter}}
 & LowPass 2kHz & 0.000$\pm$0.000 & \textbf{0.995$\pm$0.000} & 0.835$\pm$0.006 & 0.530$\pm$0.010 & 0.875$\pm$0.005 & \textbf{0.995$\pm$0.000} \\
 & HighPass 500Hz & \textbf{1.000$\pm$0.000} & 0.995$\pm$0.000 & 0.992$\pm$0.000 & 0.995$\pm$0.000 & 0.997$\pm$0.000 & 0.995$\pm$0.000 \\
\midrule
\multirow{3}{*}{\textbf{Codec}}
 & FACodec & 0.000$\pm$0.000 & 0.021$\pm$0.004 & 0.041$\pm$0.005 & 0.019$\pm$0.004 & 0.149$\pm$0.009 & \textbf{0.974$\pm$0.002} \\
 & EnCodec 6kbps$^\dagger$ & 0.000$\pm$0.000 & 0.150$\pm$0.009 & 0.072$\pm$0.007 & 0.033$\pm$0.005 & 0.252$\pm$0.010 & \textbf{0.987$\pm$0.001} \\
 & TiCodec 1g4r$^\dagger$ & 0.000$\pm$0.000 & 0.023$\pm$0.004 & 0.033$\pm$0.005 & 0.027$\pm$0.004 & 0.075$\pm$0.007 & \textbf{0.826$\pm$0.006} \\
\midrule
\multirow{4}{*}{\textbf{Denoiser}}
 & FRCRN+Denoise (0 dB)$^\dagger$ & 0.000$\pm$0.000 & 0.034$\pm$0.005 & 0.117$\pm$0.008 & 0.073$\pm$0.007 & 0.629$\pm$0.009 & \textbf{0.944$\pm$0.003} \\
 & FRCRN+Denoise (5 dB)$^\dagger$ & 0.000$\pm$0.000 & 0.079$\pm$0.007 & 0.437$\pm$0.011 & 0.224$\pm$0.010 & 0.863$\pm$0.005 & \textbf{0.991$\pm$0.001} \\
 & MossFormer+Denoise (0 dB)$^\dagger$ & 0.000$\pm$0.000 & 0.029$\pm$0.004 & 0.107$\pm$0.008 & 0.111$\pm$0.008 & 0.614$\pm$0.009 & \textbf{0.956$\pm$0.003} \\
 & MossFormer+Denoise (5 dB)$^\dagger$ & 0.000$\pm$0.000 & 0.073$\pm$0.007 & 0.368$\pm$0.011 & 0.281$\pm$0.011 & 0.860$\pm$0.005 & \textbf{0.991$\pm$0.001} \\
\midrule
\multirow{2}{*}{\textbf{Vocoder}}
 & HiFi-GAN$^\dagger$ & 0.000$\pm$0.000 & 0.014$\pm$0.003 & 0.809$\pm$0.006 & 0.302$\pm$0.010 & 0.713$\pm$0.008 & \textbf{0.995$\pm$0.000} \\
 & Vocos$^\dagger$ & 0.000$\pm$0.000 & 0.029$\pm$0.005 & 0.992$\pm$0.000 & 0.994$\pm$0.000 & \textbf{0.996$\pm$0.000} & 0.994$\pm$0.000 \\
\midrule
\multicolumn{2}{l}{\textbf{Worst Case}} & 0.000$\pm$0.000 & 0.014$\pm$0.003 & 0.033$\pm$0.005 & 0.019$\pm$0.004 & 0.015$\pm$0.003 & \textbf{0.826$\pm$0.006} \\
\bottomrule
\end{tabular}
\caption{Main robustness comparison: F1$\pm$SE across attacks, all models. $\dagger$ denotes unseen attacks.}
\label{tab:main_robustness}
\end{table*}

\begin{table}[t]
\centering
\small
\setlength{\tabcolsep}{3pt}
\begin{tabular}{l c c c}
\toprule
\textbf{Model} & \textbf{SI-SNR (dB) $\uparrow$} & \textbf{PESQ $\uparrow$} & \textbf{STOI $\uparrow$} \\
\midrule
WavMark       & \textbf{36.25$\pm$0.03} & 4.160$\pm$0.003 & 0.997$\pm$0.000 \\
AudioSeal     & 26.82$\pm$0.06 & \textbf{4.247$\pm$0.001} & \textbf{0.998$\pm$0.000} \\
TimbreWM      & 27.87$\pm$0.05 & 4.011$\pm$0.003 & 0.993$\pm$0.000 \\
WMCodec       & 1.52$\pm$0.09 & 3.168$\pm$0.007 & 0.934$\pm$0.001 \\
AWARE         & 17.34$\pm$0.08 & 4.086$\pm$0.002 & 0.985$\pm$0.000 \\
\ours (Ours)  & 18.29$\pm$0.05 & 3.990$\pm$0.003 & 0.966$\pm$0.000 \\
\bottomrule
\end{tabular}
\caption{Fidelity comparison across baselines (mean$\pm$SE).}
\label{tab:main_fidelity}
\end{table}

\paragraph{\textbf{Dataset}}
For watermark training we use the standard \texttt{train\_clean100} split of
LibriSpeech~\citep{Panayotov2015librispeech}, with the same 2620-clip test set. Audio is resampled to
22.05kHz. In Section~\ref{subsec:zeroshot} we also evaluate zero-shot
generalization to LJSpeech~\citep{Ito2017ljspeech}, without retraining.

\paragraph{\textbf{Metrics}}
For fidelity evaluation, we adopt three objective metrics:
Scale-Invariant Signal-to-Noise Ratio (SI-SNR), Perceptual Evaluation of
Speech Quality (PESQ), and
Short-Time Objective Intelligibility (STOI).  Out of these three PESQ is commonly considered the metric that best aligns with human auditory sensitivity, which is why the masking policy uses it as its criteria. \\

We report detection (whether the watermark is judged present or absent)
since that is the operative goal at inference: for every model we
calibrate a threshold on the fraction of correctly decoded bits, using
clean audio to fix a 1\% false-positive rate, then measure detection F1
under each attack at that operating point.
 Since F1 is a single aggregate number over the whole test set rather
than a per-clip quantity, we estimate its standard error via a paired
bootstrap: for each attack we resample the positive test clips with
replacement 1000 times, sharing resample indices across all models for
a valid paired comparison, and take the standard error of the
resulting F1 distribution. Fidelity metrics (SI-SNR, PESQ, STOI), by
contrast, are computed directly per clip, so we report the standard
error of the per-clip mean across the test set rather than a bootstrap
estimate. In both cases, bold marks the model(s) tied for
best, i.e.\ not significantly below the top performer (for F1, in
$\geq$95\% of resamples, $p{<}0.05$; for fidelity metrics, by more than
one combined standard error). Unless otherwise stated, crops are reported at 95\% removal,
and the full attack suite (all crop ratios and distortion levels) is
reported in the supplementary material. The default watermark length is
10 bits. The main baseline
comparison (Table~\ref{tab:main_robustness}) runs on the full 2620-clip test
set. Ablation and ECC-variant studies use a fixed 200-clip subset for
speed, as does the zero-shot evaluation on LJSpeech
(Section~\ref{subsec:zeroshot}).

\paragraph{\textbf{Distortions}}
We evaluate robustness against a broad set of distortions, split between
classical signal-processing attacks (cropping, resampling, etc.) and
neural re-synthesis, which regenerates rather than perturbs the
waveform: three neural codecs (FACodec~\citep{Ju2024naturalspeech3},
EnCodec~\citep{Defossez2023encodec}, TiCodec~\citep{Ren2024ticodec}),
two denoisers from the ClearVoice
toolkit\footnote{\url{https://github.com/modelscope/ClearerVoice-Studio}}
(FRCRN and MossFormerGAN), and two neural vocoders
(HiFi-GAN~\citep{Kong2020hifigan}, Vocos~\citep{Siuzdak2024vocos}).
FACodec is present in the training-time distortion layer and is
therefore a seen attack. TiCodec, EnCodec, HiFi-GAN, Vocos, and 
denoiser chains are held out during training.

\paragraph{\textbf{Watermarking Methods}}
We compare against five watermarking methods using their official
released implementations and checkpoints: TimbreWatermark~\citep{Liu2023timbre},
AudioSeal~\citep{SanRoman2024audioseal}, WavMark~\citep{Chen2023wavmark},
AWARE~\citep{Pavlovic2025aware}, and WMCodec~\citep{Zhou2025wmcodec}.
AudioSeal and WavMark support only a 16-bit payload, TimbreWatermark's
original checkpoint is 10-bit, we evaluate AWARE at its higher-capacity
20-bit configuration, and WMCodec uses a 16-bit payload embedded jointly
with its neural codec.

\subsection{Component Ablation}
\label{subsec:component_ablation}

Tables~\ref{tab:ablation_robustness} and~\ref{tab:ablation_fidelity} show
robustness and fidelity side by side across the same five incremental
stages: a Baseline reproducing TimbreWatermark~\citep{Liu2023timbre}, plus
the robust distortion layer (+DL), Q-Former pooling (+QFormer),
PESQ-gradient masking (+Mask), and the error-correcting code (+ECC, full
\ours). +DL alone accounts for nearly all of the robustness gain over the
Baseline but at a steep fidelity cost. The added
+QFormer improves fidelity substantially, as well as
pushes robustness past +DL on several of the hardest attacks.
+Mask recovers fidelity further still, pushing PESQ past even the
timbre baseline (4.085 vs.\ 4.018), but this comes at a cost to codec
robustness specifically: TiCodec drops from
0.948 to 0.684. We
add an error-correcting code specifically to close this gap:  +ECC recovers most of the drop across all three codecs at a
 small further cost to fidelity.

\subsection{Comparison with Existing Watermarking Methods}
\label{subsec:main}

\paragraph{\textbf{Robustness}}
We evaluate robustness under two categories of distortions: neural
re-synthesis (codecs, denoisers, and vocoders) and traditional
signal-level distortions (cropping, resampling, filtering, and the
like). Table~\ref{tab:main_robustness} shows F1 across every attack.
Under neural re-synthesis, \ours beats every baseline on nearly
every attack, most decisively on codecs, where all five baselines
degrade substantially, collapsing toward chance in several cases, and
even the strongest baseline never exceeds an F1 score of 0.252. Under
traditional signal-level distortions, \ours is comparable to prior
work in most conditions, with clear improvements under MP3 compression
and median filtering.

\paragraph{\textbf{Fidelity}}
PESQ is the metric we treat as the primary fidelity criterion, since
it best predicts what a listener perceives and is what \ours's masking
policy directly optimizes against. \ours reaches a PESQ of 3.99,
comparable to the baselines (3.17--4.25) and well above the weakest,
WMCodec (3.17); scores above 4 are considered perceptually near-transparent. SI-SNR and STOI follow the same pattern: good scores that show we maintain high fidelity, though lower than the best results.

Overall, \ours offers a good trade-off between robustness and fidelity: it offers significantly stronger robustness results while maintaining fidelity at levels that are imperceptible to the casual listener.\footnote{Audio samples are available at \url{https://davidc1212.github.io/craw-audio-samples/}.}


\subsection{Ablation Studies}
\label{subsec:ablation}

\paragraph{\textbf{Masking Threshold}}
\label{para:masking}
We sweep the masking ratio $k$ from 0\% to 100\% and measure
robustness and fidelity at each point. We find that robustness degrades slowly as $k$ increases while fidelity increases sharply. We set  $k{=}10\%$ as our operating point, as it gives a good tradeoff between fidelity and robustness. We also compare our PESQ-gradient mask with random masks at equivalent k values and show that it recovers substantially more fidelity throughout the sweep. (see supplementary material).


\paragraph{\textbf{Error-Correcting Code}}
\label{para:ecc}
To evaluate the error-correcting code, we test a few variants: a
repetition code, Reed--Solomon, LDPC, and no coding at all, all
with the same $k{=}10\%$ masking applied (per-attack robustness and
fidelity tables in the supplementary material). Rep3 is the simplest
of the four, but surprisingly it wins against every alternative and improves
robustness on codecs over no coding at all. We note that codec
corruption spreads bit errors roughly uniformly across the codeword
rather than concentrating them in bursts, exactly the pattern
repetition coding is built to correct, which is why Rep3 outperforms
the more structured alternatives specifically there. Fidelity is
comparable across all variants, so the ECC costs little on that
front. 


\subsection{Cross-Domain Generalization}
\label{subsec:zeroshot}
We evaluate how \ours generalizes to a different dataset within the
same speech domain. LJSpeech is a single-speaker corpus recorded under
different conditions than LibriSpeech; without any retraining, \ours
transfers zero-shot to this new dataset (see full per-attack tables in the
supplementary material). Codec F1 score mostly transfers or improves,
with a modest decline on TiCodec (0.838$\to$0.794); PESQ is unchanged
at 4.005 in both domains. This shows that \ours can generalize to other speech datasets without fine-tuning. 

\subsection{\textbf{Limitations}} PESQ-gradient masking is computed once
per clip at inference time and is not free: it adds roughly
150\,ms of GPU latency per clip on top of the base embedder, an $8\times$ increase over encoder-only inference (21\,ms vs.\ 171\,ms, measured over 100 clips ranging 1.82--10.52\,s, see supplementary material). While a significant increase in relative terms, 171\,ms is not a significant overhead for non-real-time applications. For real-time application, this would depend on the RTF of the generation method. Reducing this overhead would be the focus of future work.

\section{Conclusion}
\label{sec:conclusion}

This paper introduced \ours, a codec-robust audio watermarking
framework built on TimbreWatermark to close its remaining robustness
gap to neural re-synthesis, specifically neural codecs.
\ours combines distortion-aware training with an attention-based
pooling mechanism, inference-time perceptual masking, and an
error-correcting code to recover the fidelity lost during robust
training. As a result, \ours achieves state-of-the-art robustness
against neural re-synthesis, most notably codec attacks that collapse
current methods, while maintaining fidelity that is comparable to other baselines.



{\small\bibliography{references}}

\clearpage
\renewcommand{\thefigure}{S\arabic{figure}}
\renewcommand{\thetable}{S\arabic{table}}
\renewcommand{\thesection}{S\arabic{section}}
\setcounter{figure}{0}
\setcounter{table}{0}
\setcounter{section}{0}

\section{Implementation Details and Hyperparameters}
\label{sec:supp_impl}
\ours reuses TimbreWatermark's convolutional encoder--embedder--extractor
architecture (skip-gated convolutional blocks, hidden dimension 64: 3
layers for the carrier encoder, 4 for the embedder, 6 for the extractor)
and its STFT front end (1024-point FFT, 256-sample hop, Hann window,
audio resampled to 22.05kHz), apart from the Q-Former pooling change
described above. The Q-Former replaces uniform frame
averaging with a single learnable query attending over all frames
through one multi-head cross-attention block (3 heads, feed-forward
expansion $4\times$, dropout 0.1) followed by a feed-forward block, each
with a pre-norm residual connection. The 10-bit message is expanded by
Rep3 to a 30-bit codeword before embedding. We train on a single GPU
with Adam ($\text{lr}=2\times10^{-5}$, $\beta_1=0.9$, $\beta_2=0.98$,
$\epsilon=10^{-9}$, no weight decay, gradients clipped to norm 1.0; a
separate Adam instance with identical settings trains the
discriminator), decayed via StepLR (step size 5000, $\gamma=0.98$), for
20 epochs at batch size 1. The training objective weights the fidelity
loss at 1.0, the message loss at 10.0 on distorted audio and 0.01 on
clean audio, and the adversarial term at 0.01. The PESQ-gradient masking
threshold ($k{=}10\%$) is applied only at inference
time and does not affect training. All models are trained with a single
fixed random seed (2022), covering Python, NumPy, and PyTorch/CUDA RNGs.
Experiments are run on a single NVIDIA L4 GPU (22.5GB) with an Intel Xeon
Gold 6530 CPU, on Rocky Linux 9.8, using PyTorch 2.1.2 (CUDA 12.1).

\section{Distortion and Attack Details}
\label{sec:supp_distortions}
We elaborate below on every audio editing augmentation used in this
work, covering both the stochastic pool applied during training and
the full attack suite applied during evaluation.

\begin{itemize}
\item \textbf{Identity.} No distortion is applied. Included in the
training pool so the model also sees clean audio and is not forced to
trade off fidelity against robustness on every single step.
\item \textbf{Additive noise.} During training, Gaussian noise is
added directly to the waveform with a standard deviation sampled
uniformly between 0.01 and 0.08 each step. At evaluation, noise is
instead added at a fixed target SNR rather than a fixed standard
deviation, swept from 20 to 40\,dB.
\item \textbf{Low-pass filter.} During training, the cutoff frequency
is re-sampled uniformly between 2 and 8\,kHz every step, so the model
sees a range of severities rather than one fixed filter. At
evaluation this is replaced by a single fixed cutoff of 2\,kHz.
\item \textbf{High-pass filter.} Evaluation only, at a fixed cutoff of
500\,Hz; there is no high-pass filter in the training pool.
\item \textbf{Amplitude scaling.} During training, the waveform is
multiplied by a factor sampled between 0.8 and 1.2, i.e.\ mild
loudness changes in either direction. At evaluation, the waveform is
instead only ever scaled down, swept from 20\% to 80\% of its
original amplitude.
\item \textbf{Mel-spectrogram resynthesis.} Training only: the
waveform is converted to a mel-spectrogram and reconstructed with the
Griffin-Lim algorithm, simulating the lossy analysis-synthesis step
common to vocoder-based resynthesis pipelines.
\item \textbf{Spectral gating.} Training only: frequency bins more
than a threshold, sampled between 10 and 30\,dB, below the frame's
peak energy are zeroed out. This acts as a differentiable proxy for a
denoiser stripping out the quiet regions of the spectrum where
watermarks often hide.
\item \textbf{FACodec resynthesis.} The watermarked waveform is
encoded, quantized, and decoded through the NaturalSpeech3 FACodec
neural codec. During training this attack is drawn 10\% of the time
(the other six training-time attacks above share the remaining 15\%
each). It is the only codec seen during training, so at evaluation it
is also the only codec attack \ours is scored on as a seen rather
than an unseen attack.
\item \textbf{Crop front / middle / back.} Evaluation only: a
contiguous chunk of the waveform is removed from its start, middle,
or end, swept from 20\% to 80\% removed in coarse steps and then
densely from 90\% to 98\% removed, where robustness begins to
degrade.
\item \textbf{Resample.} Evaluation only: the waveform is
downsampled to 16\,kHz or to 8\,kHz and then upsampled back to the
original rate, introducing the quantization and aliasing artifacts of
the round trip.
\item \textbf{MP3 compression.} Evaluation only: the waveform is
re-encoded through an MP3 codec at a fixed bitrate, swept from 8 to
64\,kbps.
\item \textbf{Recount 8bps.} Evaluation only: the waveform is
requantized to 8 bits per sample, a fixed, single-severity attack.
\item \textbf{Median filtering.} Evaluation only: a median filter
slides over the waveform, evaluated at window sizes of 5, 15, 25, and
35 samples.
\item \textbf{Time stretch.} Evaluation only: playback speed is
changed without altering pitch, swept from $-20\%$ to $20\%$ (i.e.\
up to 20\% slower or 20\% faster).
\item \textbf{Denoising.} Evaluation only: Gaussian noise is added at
a target SNR and then removed with a neural denoiser, either
ClearVoice FRCRN or ClearVoice MossFormerGAN, to test whether the
denoiser strips the watermark out along with the noise, swept from 0
to 20\,dB SNR.
\item \textbf{Unseen neural codecs.} Evaluation only: TiCodec and
EnCodec, neither of which the model encounters during training.
\item \textbf{Neural vocoders.} Evaluation only: HiFi-GAN and Vocos,
both re-synthesizing the waveform through a mel-spectrogram
intermediate; neither is seen during training.
\end{itemize}

\section{Masking Threshold Sweep}
\label{sec:supp_masking}
\begin{figure}[h]
\centering
\includegraphics[width=\linewidth]{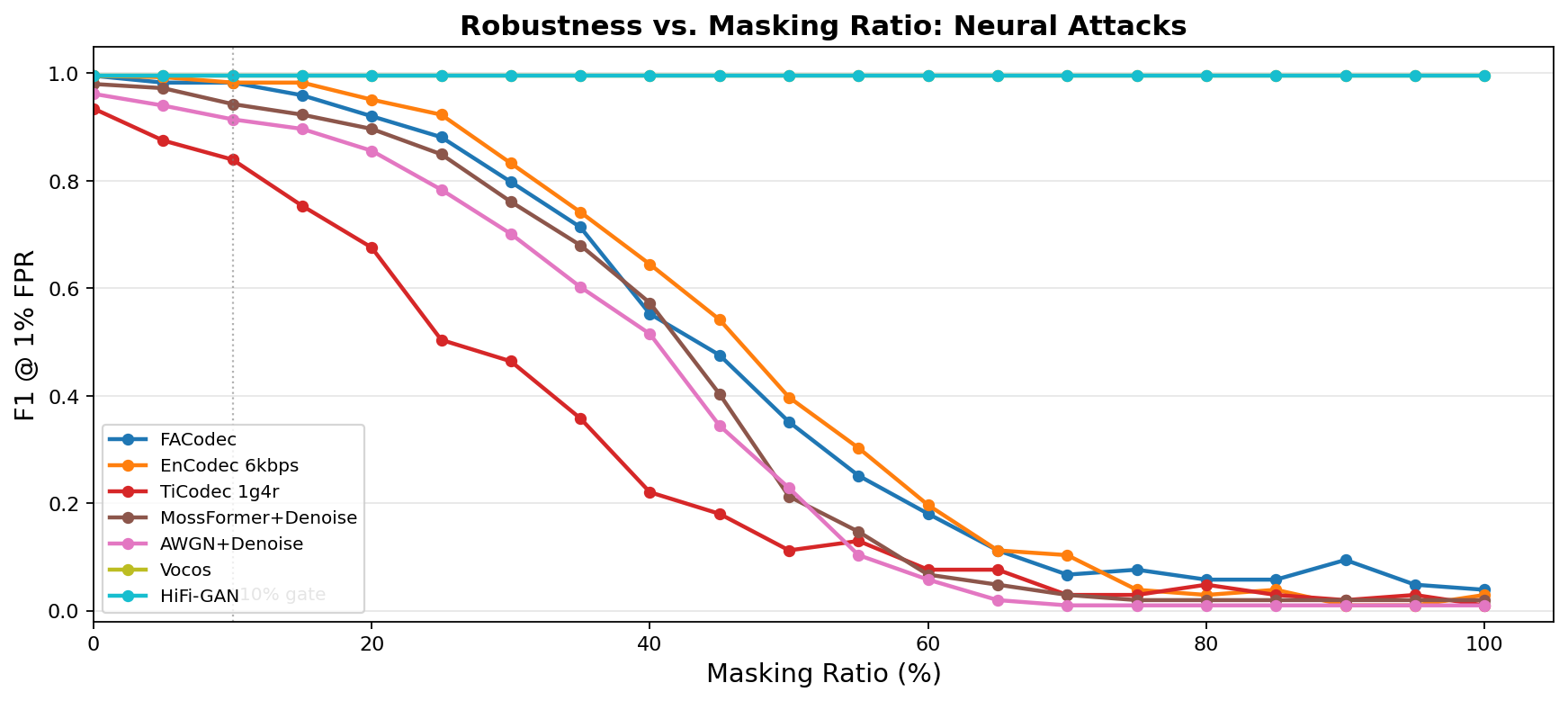}
\caption{Robustness (F1@1\%FPR) vs.\ masking ratio under neural attacks. Robustness degrades steadily with masking.}
\label{fig:supp_ablation_neural}
\end{figure}

\begin{figure}[h]
\centering
\includegraphics[width=\linewidth]{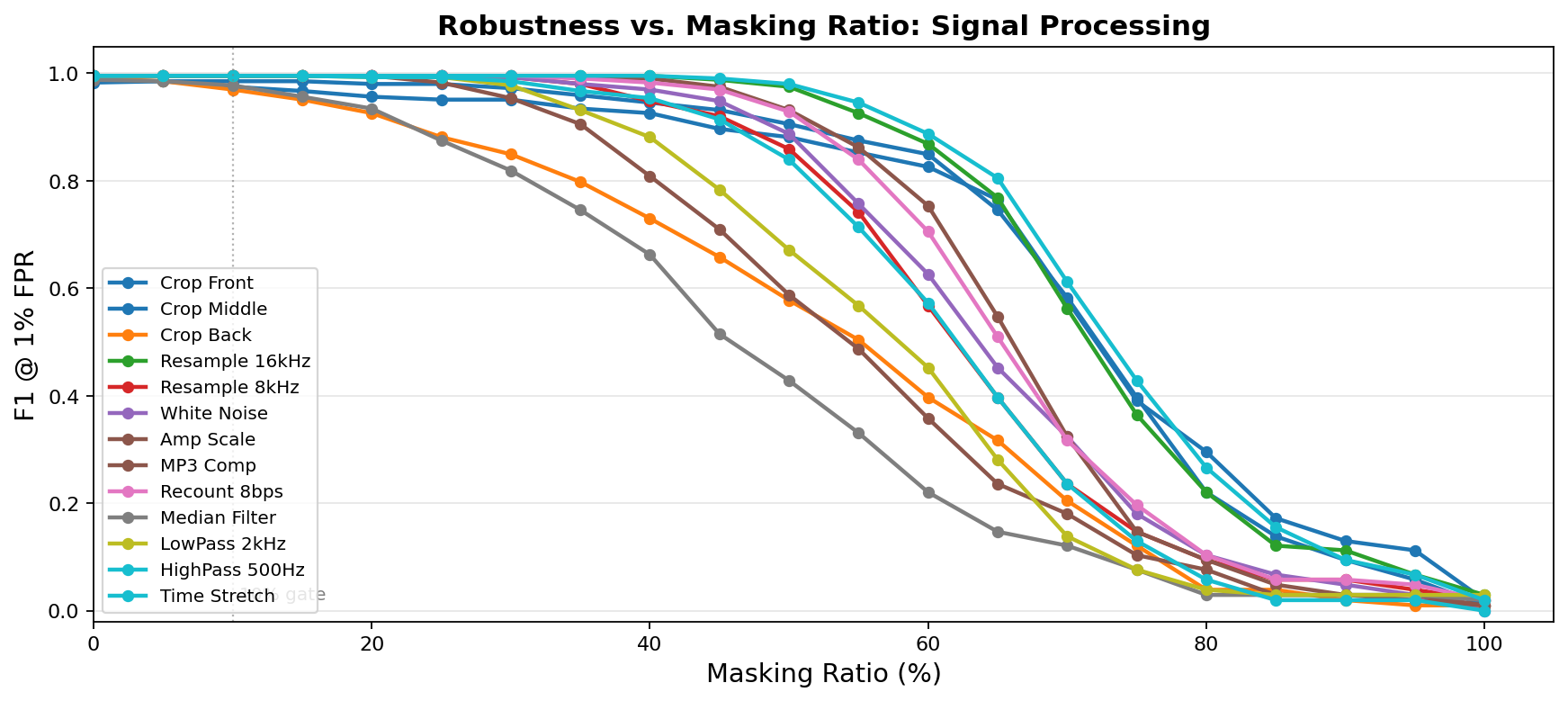}
\caption{Robustness (F1@1\%FPR) vs.\ masking ratio under signal-processing attacks.}
\label{fig:supp_ablation_sigproc}
\end{figure}

\begin{figure}[h]
\centering
\includegraphics[width=\columnwidth]{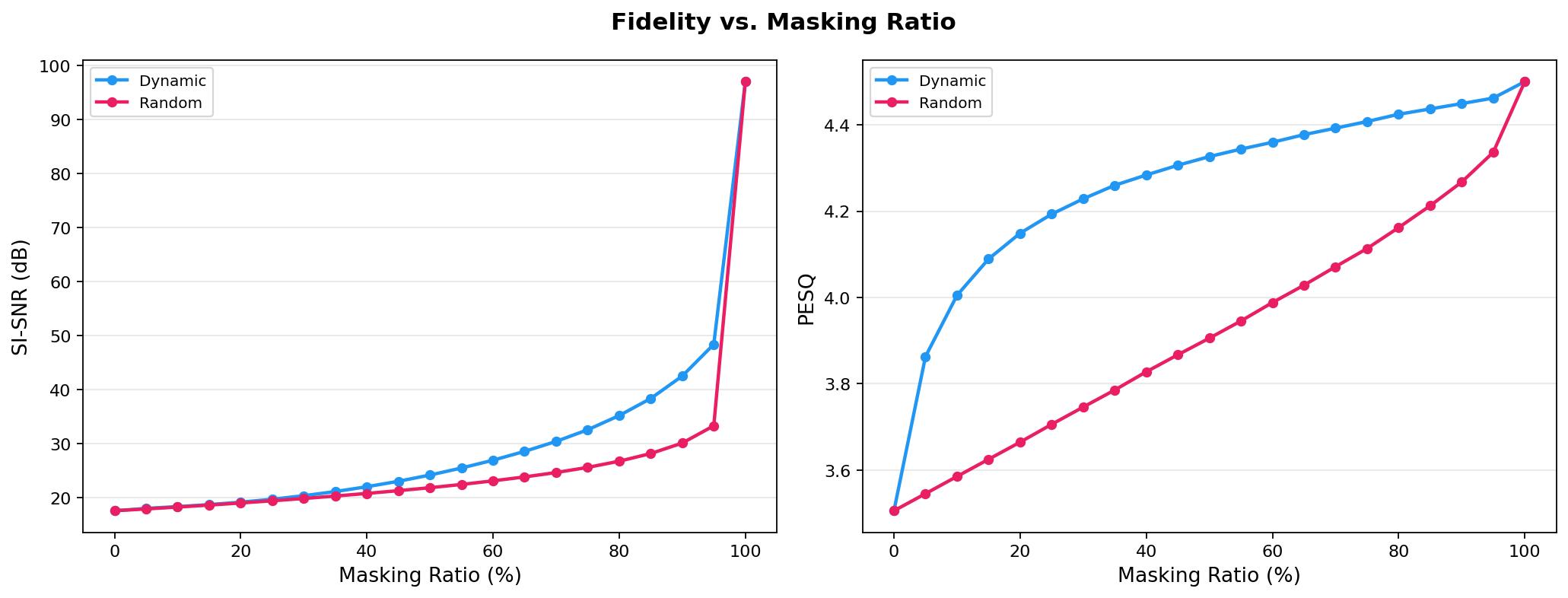}
\caption{Fidelity as a function of the PESQ-gradient masking threshold,
comparing dynamic masking (top-$k\%$ gated by PESQ gradient) against random masking.
The PESQ-gradient mask recovers substantially more fidelity than random
masks at equivalent $k$.}
\label{fig:supp_masking_fidelity}
\end{figure}

Figures~\ref{fig:supp_ablation_neural} and~\ref{fig:supp_ablation_sigproc}
show robustness as a function of the masking ratio $k$ under neural and
signal-processing attacks respectively, with the error-correcting code
(Rep3) enabled throughout; only $k$ varies.
Figure~\ref{fig:supp_masking_fidelity} shows the corresponding fidelity
sweep, comparing the PESQ-gradient mask against a random mask of the same
size at equivalent $k$.

\FloatBarrier

\section{Masking Overhead}
\label{sec:supp_masking_overhead}
Table~\ref{tab:masking_latency} reports the per-clip GPU inference
latency added by PESQ-gradient dynamic masking, referenced in the
Limitations subsection.

\bigskip
\begingroup
\centering
\small
\begin{tabular}{l c}
\toprule
\textbf{Condition} & \textbf{Latency (ms) $\downarrow$} \\
\midrule
Baseline (encoder only)        & 21.20$\pm$4.53 \\
Full pipeline (encoder + mask) & 170.93$\pm$7.83 \\
\bottomrule
\end{tabular}
\captionof{table}{Per-clip GPU inference latency (mean$\pm$SE, $n{=}100$
LibriSpeech test clips, durations 1.82--10.52\,s) for the embedder
alone vs.\ the full pipeline with PESQ-gradient dynamic masking
(10\% masking). Masking adds $\sim$150\,ms/clip.}
\label{tab:masking_latency}
\endgroup

\FloatBarrier

\section{Component Ablation: Full Ratio Sweep}
\label{sec:supp_ablation_robustness}
Tables~\ref{tab:supp_ablation_robustness_full_sigproc}
and~\ref{tab:supp_ablation_robustness_full_neural} report the full
ratio sweep underlying the component ablation above, split
by traditional signal-level distortions and neural re-synthesis
attacks respectively, across all five incremental stages
(Baseline$\to$+DL$\to$+QFormer$\to$+Mask$\to$+ECC).

\section{Baseline Comparison: Full Ratio Sweep}
\label{sec:supp_main_robustness}
Tables~\ref{tab:supp_main_robustness_full_sigproc}
and~\ref{tab:supp_main_robustness_full_neural} report the full ratio
sweep underlying the main robustness comparison, split by
signal-level and neural re-synthesis attacks.

\section{Error-Correcting Code Variants}
\label{sec:supp_ecc}
For each error-correcting code variant, we train a separate model with
that ECC wired into the embedding and extraction pipeline, apply
dynamic PESQ masking ($k{=}10\%$) at inference time, and evaluate
robustness and fidelity under the full attack suite (fidelity reported
above in Table~\ref{tab:supp_ecc_fidelity}). We compare five
variants: Rep3, Reed--Solomon (RS(4,2), RS(6,2)), LDPC, and no coding,
referenced in the Error-Correcting Code paragraph of Ablation Studies
above. RS(4,2) and RS(6,2) operate over GF($2^8$) with
$n{-}k{=}2$ and $4$ parity bytes respectively, correcting up to 1 and 2
byte errors per codeword (32 and 48 codeword bits for a 16-bit
message). LDPC uses a regular Gallager
(1962) construction at rate $1/2$ ($n{=}20$, $k{=}10$, variable-node
degree $d_v{=}3$, check-node degree $d_c{=}6$), decoded with min-sum
belief propagation. Tables~\ref{tab:supp_ecc_robustness_full_sigproc}
and~\ref{tab:supp_ecc_robustness_full_neural} report the full ratio
sweep, split by signal-level and neural re-synthesis attacks.

\bigskip

\begingroup
\centering
\footnotesize
\setlength{\tabcolsep}{3pt}
\begin{tabular}{l c c c}
\toprule
\textbf{Variant} & \textbf{SI-SNR (dB) $\uparrow$} & \textbf{PESQ $\uparrow$} & \textbf{STOI $\uparrow$} \\
\midrule
No ECC   & \textbf{19.12$\pm$0.20} & \textbf{4.085$\pm$0.008} & \textbf{0.974$\pm$0.001} \\
Rep3     & 18.39$\pm$0.17 & 4.005$\pm$0.011 & 0.966$\pm$0.001 \\
RS(4,2)  & 14.63$\pm$0.18 & 3.876$\pm$0.013 & 0.952$\pm$0.001 \\
RS(6,2)  & 14.32$\pm$0.18 & 4.033$\pm$0.011 & 0.960$\pm$0.001 \\
LDPC     & 15.92$\pm$0.17 & 4.069$\pm$0.007 & 0.963$\pm$0.001 \\
\bottomrule
\end{tabular}
\captionof{table}{ECC variant fidelity (mean$\pm$SE) with dynamic PESQ masking (10\%), 200-clip subset.}
\label{tab:supp_ecc_fidelity}
\endgroup

\FloatBarrier

\section{Cross-Domain Generalization: LJSpeech}
\label{sec:supp_zeroshot}
Tables~\ref{tab:supp_zeroshot_full_sigproc} and~\ref{tab:supp_zeroshot_full_neural}
report the full ratio sweep of per-attack robustness, in-domain
(LibriSpeech) vs.\ zero-shot (LJSpeech), split by signal-level and
neural re-synthesis attacks (corresponding fidelity reported above in
Table~\ref{tab:supp_zeroshot_fidelity}).

\bigskip
\begingroup
\centering
\footnotesize
\setlength{\tabcolsep}{1pt}
\begin{tabular}{l c c c}
\toprule
\textbf{Model} & \textbf{SI-SNR (dB) $\uparrow$} & \textbf{PESQ $\uparrow$} & \textbf{STOI $\uparrow$} \\
\midrule
\ours (LibriSpeech)     & 18.39$\pm$0.17 & 4.005$\pm$0.011 & 0.966$\pm$0.001 \\
Zero-shot (LJSpeech)    & \textbf{19.40$\pm$0.07} & 4.005$\pm$0.005 & \textbf{0.976$\pm$0.000} \\
\bottomrule
\end{tabular}
\captionof{table}{Fidelity comparison (mean$\pm$SE): \ours in-domain (LibriSpeech) vs.\ zero-shot on LJSpeech, 200-clip subset per domain.}
\label{tab:supp_zeroshot_fidelity}
\endgroup

\FloatBarrier

\begin{table*}[t]
\centering
\small
\setlength{\tabcolsep}{4pt}
\begin{tabular}{l l c c c c c}
\toprule
\textbf{Category} & \textbf{Attack} & \textbf{Baseline} & \textbf{+DL} & \textbf{+QFormer} & \textbf{+Mask} & \textbf{+ECC} \\
\midrule
\multirow{57}{*}{\textbf{Sig.-Proc.}}
 & Crop Front (20\%) & \textbf{0.998$\pm$0.000} & 0.994$\pm$0.000 & 0.995$\pm$0.000 & 0.996$\pm$0.000 & 0.994$\pm$0.000 \\
 & Crop Front (40\%) & \textbf{0.998$\pm$0.000} & 0.994$\pm$0.000 & 0.993$\pm$0.002 & 0.993$\pm$0.002 & 0.994$\pm$0.000 \\
 & Crop Front (60\%) & \textbf{0.998$\pm$0.000} & 0.994$\pm$0.000 & 0.993$\pm$0.002 & 0.993$\pm$0.002 & 0.994$\pm$0.000 \\
 & Crop Front (80\%) & \textbf{0.998$\pm$0.000} & 0.994$\pm$0.000 & 0.980$\pm$0.006 & 0.980$\pm$0.006 & 0.994$\pm$0.000 \\
 & Crop Front (90\%) & \textbf{0.998$\pm$0.000} & 0.994$\pm$0.000 & 0.959$\pm$0.010 & 0.959$\pm$0.010 & 0.994$\pm$0.000 \\
 & Crop Front (95\%) & \textbf{0.998$\pm$0.000} & 0.994$\pm$0.000 & 0.964$\pm$0.009 & 0.965$\pm$0.009 & 0.994$\pm$0.000 \\
 & Crop Front (96\%) & \textbf{0.998$\pm$0.000} & 0.992$\pm$0.002 & 0.961$\pm$0.009 & 0.962$\pm$0.009 & 0.994$\pm$0.000 \\
 & Crop Front (97\%) & \textbf{0.995$\pm$0.003} & 0.986$\pm$0.004 & 0.982$\pm$0.006 & 0.983$\pm$0.006 & \textbf{0.994$\pm$0.000} \\
 & Crop Front (98\%) & 0.972$\pm$0.008 & \textbf{0.974$\pm$0.007} & 0.982$\pm$0.006 & \textbf{0.985$\pm$0.005} & \textbf{0.984$\pm$0.005} \\
 & Crop Middle (20\%) & \textbf{0.998$\pm$0.000} & 0.994$\pm$0.000 & 0.995$\pm$0.000 & 0.996$\pm$0.000 & 0.994$\pm$0.000 \\
 & Crop Middle (40\%) & \textbf{0.998$\pm$0.000} & 0.994$\pm$0.000 & 0.995$\pm$0.000 & 0.996$\pm$0.000 & 0.994$\pm$0.000 \\
 & Crop Middle (60\%) & \textbf{0.998$\pm$0.000} & 0.994$\pm$0.000 & 0.995$\pm$0.000 & 0.996$\pm$0.000 & 0.994$\pm$0.000 \\
 & Crop Middle (80\%) & \textbf{0.998$\pm$0.000} & 0.994$\pm$0.000 & 0.995$\pm$0.000 & 0.996$\pm$0.000 & 0.994$\pm$0.000 \\
 & Crop Middle (90\%) & \textbf{0.998$\pm$0.000} & 0.994$\pm$0.000 & 0.995$\pm$0.000 & 0.996$\pm$0.000 & 0.994$\pm$0.000 \\
 & Crop Middle (95\%) & \textbf{0.998$\pm$0.000} & 0.994$\pm$0.000 & 0.995$\pm$0.000 & 0.996$\pm$0.000 & 0.992$\pm$0.003 \\
 & Crop Middle (96\%) & \textbf{0.998$\pm$0.000} & 0.994$\pm$0.000 & 0.995$\pm$0.000 & 0.996$\pm$0.000 & 0.994$\pm$0.000 \\
 & Crop Middle (97\%) & \textbf{0.995$\pm$0.003} & 0.986$\pm$0.004 & 0.995$\pm$0.000 & \textbf{0.996$\pm$0.000} & 0.987$\pm$0.004 \\
 & Crop Middle (98\%) & \textbf{0.995$\pm$0.002} & 0.952$\pm$0.010 & \textbf{0.993$\pm$0.002} & \textbf{0.993$\pm$0.002} & 0.974$\pm$0.007 \\
 & Crop Back (20\%) & \textbf{0.998$\pm$0.000} & 0.994$\pm$0.000 & 0.995$\pm$0.000 & 0.996$\pm$0.000 & 0.994$\pm$0.000 \\
 & Crop Back (40\%) & \textbf{0.998$\pm$0.000} & 0.994$\pm$0.000 & 0.995$\pm$0.000 & 0.996$\pm$0.000 & 0.994$\pm$0.000 \\
 & Crop Back (60\%) & \textbf{0.998$\pm$0.000} & 0.994$\pm$0.000 & 0.995$\pm$0.000 & 0.996$\pm$0.000 & 0.994$\pm$0.000 \\
 & Crop Back (80\%) & \textbf{0.998$\pm$0.000} & 0.994$\pm$0.000 & 0.995$\pm$0.000 & 0.996$\pm$0.000 & 0.994$\pm$0.000 \\
 & Crop Back (90\%) & \textbf{0.998$\pm$0.000} & 0.994$\pm$0.000 & 0.993$\pm$0.003 & 0.993$\pm$0.003 & 0.992$\pm$0.003 \\
 & Crop Back (95\%) & \textbf{0.998$\pm$0.000} & 0.994$\pm$0.000 & 0.993$\pm$0.003 & 0.993$\pm$0.003 & 0.992$\pm$0.003 \\
 & Crop Back (96\%) & \textbf{0.998$\pm$0.000} & 0.994$\pm$0.000 & 0.990$\pm$0.003 & 0.990$\pm$0.003 & 0.984$\pm$0.005 \\
 & Crop Back (97\%) & \textbf{0.998$\pm$0.000} & 0.989$\pm$0.004 & 0.987$\pm$0.004 & 0.988$\pm$0.004 & 0.974$\pm$0.007 \\
 & Crop Back (98\%) & \textbf{0.998$\pm$0.000} & 0.981$\pm$0.006 & 0.985$\pm$0.005 & 0.985$\pm$0.005 & 0.969$\pm$0.008 \\
 & Resample 16kHz & \textbf{0.998$\pm$0.000} & 0.994$\pm$0.000 & 0.995$\pm$0.000 & 0.996$\pm$0.000 & 0.994$\pm$0.000 \\
 & Resample 8kHz & \textbf{0.995$\pm$0.002} & 0.994$\pm$0.000 & 0.995$\pm$0.000 & \textbf{0.996$\pm$0.000} & 0.994$\pm$0.000 \\
 & White Noise (20 dB) & \textbf{0.998$\pm$0.000} & 0.994$\pm$0.000 & 0.995$\pm$0.000 & 0.996$\pm$0.000 & 0.994$\pm$0.000 \\
 & White Noise (25 dB) & \textbf{0.998$\pm$0.000} & 0.994$\pm$0.000 & 0.995$\pm$0.000 & 0.996$\pm$0.000 & 0.994$\pm$0.000 \\
 & White Noise (30 dB) & \textbf{0.998$\pm$0.000} & 0.994$\pm$0.000 & 0.995$\pm$0.000 & 0.996$\pm$0.000 & 0.994$\pm$0.000 \\
 & White Noise (35 dB) & \textbf{0.998$\pm$0.000} & 0.994$\pm$0.000 & 0.995$\pm$0.000 & 0.996$\pm$0.000 & 0.994$\pm$0.000 \\
 & White Noise (40 dB) & \textbf{0.998$\pm$0.000} & 0.994$\pm$0.000 & 0.995$\pm$0.000 & 0.996$\pm$0.000 & 0.994$\pm$0.000 \\
 & Amp Scale (20\%) & \textbf{0.998$\pm$0.000} & 0.994$\pm$0.000 & 0.995$\pm$0.000 & 0.996$\pm$0.000 & 0.994$\pm$0.000 \\
 & Amp Scale (40\%) & \textbf{0.998$\pm$0.000} & 0.994$\pm$0.000 & 0.995$\pm$0.000 & 0.996$\pm$0.000 & 0.994$\pm$0.000 \\
 & Amp Scale (60\%) & \textbf{0.998$\pm$0.000} & 0.994$\pm$0.000 & 0.995$\pm$0.000 & 0.996$\pm$0.000 & 0.994$\pm$0.000 \\
 & Amp Scale (80\%) & \textbf{0.998$\pm$0.000} & 0.994$\pm$0.000 & 0.995$\pm$0.000 & 0.996$\pm$0.000 & 0.994$\pm$0.000 \\
 & MP3 Comp (8 kbps) & 0.861$\pm$0.020 & 0.994$\pm$0.000 & 0.995$\pm$0.000 & \textbf{0.996$\pm$0.000} & 0.994$\pm$0.000 \\
 & MP3 Comp (16 kbps) & \textbf{0.998$\pm$0.000} & 0.994$\pm$0.000 & 0.995$\pm$0.000 & 0.996$\pm$0.000 & 0.994$\pm$0.000 \\
 & MP3 Comp (24 kbps) & \textbf{0.998$\pm$0.000} & 0.994$\pm$0.000 & 0.995$\pm$0.000 & 0.996$\pm$0.000 & 0.994$\pm$0.000 \\
 & MP3 Comp (32 kbps) & \textbf{0.998$\pm$0.000} & 0.994$\pm$0.000 & 0.995$\pm$0.000 & 0.996$\pm$0.000 & 0.994$\pm$0.000 \\
 & MP3 Comp (40 kbps) & \textbf{0.998$\pm$0.000} & 0.994$\pm$0.000 & 0.995$\pm$0.000 & 0.996$\pm$0.000 & 0.994$\pm$0.000 \\
 & MP3 Comp (48 kbps) & \textbf{0.998$\pm$0.000} & 0.994$\pm$0.000 & 0.995$\pm$0.000 & 0.996$\pm$0.000 & 0.994$\pm$0.000 \\
 & MP3 Comp (56 kbps) & \textbf{0.998$\pm$0.000} & 0.994$\pm$0.000 & 0.995$\pm$0.000 & 0.996$\pm$0.000 & 0.994$\pm$0.000 \\
 & MP3 Comp (64 kbps) & \textbf{0.998$\pm$0.000} & 0.994$\pm$0.000 & 0.995$\pm$0.000 & 0.996$\pm$0.000 & 0.994$\pm$0.000 \\
 & Recount 8bps & \textbf{0.998$\pm$0.000} & 0.994$\pm$0.000 & 0.995$\pm$0.000 & 0.996$\pm$0.000 & 0.994$\pm$0.000 \\
 & Median Filt (5) & \textbf{0.998$\pm$0.000} & 0.994$\pm$0.000 & 0.995$\pm$0.000 & 0.996$\pm$0.000 & 0.994$\pm$0.000 \\
 & Median Filt (15) & \textbf{0.995$\pm$0.003} & 0.992$\pm$0.002 & 0.995$\pm$0.000 & \textbf{0.996$\pm$0.000} & 0.994$\pm$0.000 \\
 & Median Filt (25) & 0.974$\pm$0.008 & 0.950$\pm$0.011 & 0.995$\pm$0.000 & \textbf{0.996$\pm$0.000} & 0.989$\pm$0.003 \\
 & Median Filt (35) & 0.893$\pm$0.016 & 0.855$\pm$0.021 & 0.982$\pm$0.005 & \textbf{0.983$\pm$0.005} & \textbf{0.976$\pm$0.007} \\
 & Time Stretch (-20\%) & \textbf{0.998$\pm$0.000} & 0.994$\pm$0.000 & 0.995$\pm$0.000 & 0.996$\pm$0.000 & 0.994$\pm$0.000 \\
 & Time Stretch (-10\%) & \textbf{0.998$\pm$0.000} & 0.994$\pm$0.000 & 0.995$\pm$0.000 & 0.996$\pm$0.000 & 0.994$\pm$0.000 \\
 & Time Stretch (-5\%) & \textbf{0.998$\pm$0.000} & 0.994$\pm$0.000 & 0.995$\pm$0.000 & 0.996$\pm$0.000 & 0.994$\pm$0.000 \\
 & Time Stretch (5\%) & \textbf{0.998$\pm$0.000} & 0.994$\pm$0.000 & 0.995$\pm$0.000 & 0.996$\pm$0.000 & 0.994$\pm$0.000 \\
 & Time Stretch (10\%) & \textbf{0.998$\pm$0.000} & 0.994$\pm$0.000 & 0.995$\pm$0.000 & 0.996$\pm$0.000 & 0.994$\pm$0.000 \\
 & Time Stretch (20\%) & \textbf{0.998$\pm$0.000} & 0.994$\pm$0.000 & 0.995$\pm$0.000 & 0.996$\pm$0.000 & 0.994$\pm$0.000 \\
\midrule
\multirow{2}{*}{\textbf{Filter}}
 & LowPass 2kHz & 0.855$\pm$0.020 & 0.994$\pm$0.000 & 0.995$\pm$0.000 & \textbf{0.996$\pm$0.000} & 0.994$\pm$0.000 \\
 & HighPass 500Hz & \textbf{0.998$\pm$0.000} & 0.979$\pm$0.006 & 0.995$\pm$0.000 & 0.996$\pm$0.000 & 0.994$\pm$0.000 \\
\bottomrule
\end{tabular}
\caption{Full ratio sweep of the component ablation under traditional signal-level distortions (F1$\pm$SE at each incremental stage, 200-clip subset). Supplementary version of the component ablation table above, part 1 of 2.}
\label{tab:supp_ablation_robustness_full_sigproc}
\end{table*}

\FloatBarrier

\begin{table*}[t]
\centering
\small
\setlength{\tabcolsep}{4pt}
\begin{tabular}{l l c c c c c}
\toprule
\textbf{Category} & \textbf{Attack} & \textbf{Baseline} & \textbf{+DL} & \textbf{+QFormer} & \textbf{+Mask} & \textbf{+ECC} \\
\midrule
\multirow{3}{*}{\textbf{Codec}}
 & FACodec & 0.029$\pm$0.016 & \textbf{0.994$\pm$0.000} & 0.934$\pm$0.013 & 0.937$\pm$0.012 & 0.982$\pm$0.005 \\
 & EnCodec 6kbps$^\dagger$ & 0.077$\pm$0.026 & \textbf{0.994$\pm$0.000} & \textbf{0.993$\pm$0.003} & 0.926$\pm$0.014 & 0.982$\pm$0.006 \\
 & TiCodec 1g4r$^\dagger$ & 0.029$\pm$0.016 & 0.895$\pm$0.016 & \textbf{0.948$\pm$0.011} & 0.684$\pm$0.030 & 0.838$\pm$0.021 \\
\midrule
\multirow{10}{*}{\textbf{Denoiser}}
 & FRCRN+Denoise (0 dB)$^\dagger$ & 0.067$\pm$0.024 & 0.818$\pm$0.022 & 0.942$\pm$0.012 & \textbf{0.943$\pm$0.012} & 0.913$\pm$0.015 \\
 & FRCRN+Denoise (5 dB)$^\dagger$ & 0.372$\pm$0.040 & \textbf{0.981$\pm$0.006} & \textbf{0.987$\pm$0.004} & \textbf{0.988$\pm$0.004} & \textbf{0.989$\pm$0.003} \\
 & FRCRN+Denoise (10 dB)$^\dagger$ & 0.800$\pm$0.023 & 0.994$\pm$0.000 & 0.995$\pm$0.000 & \textbf{0.996$\pm$0.000} & 0.992$\pm$0.003 \\
 & FRCRN+Denoise (15 dB)$^\dagger$ & 0.969$\pm$0.008 & 0.994$\pm$0.000 & 0.995$\pm$0.000 & \textbf{0.996$\pm$0.000} & 0.994$\pm$0.000 \\
 & FRCRN+Denoise (20 dB)$^\dagger$ & \textbf{0.992$\pm$0.004} & 0.994$\pm$0.000 & 0.995$\pm$0.000 & \textbf{0.996$\pm$0.000} & 0.994$\pm$0.000 \\
 & MossFormer+Denoise (0 dB)$^\dagger$ & 0.104$\pm$0.027 & \textbf{0.963$\pm$0.009} & \textbf{0.953$\pm$0.010} & \textbf{0.954$\pm$0.010} & \textbf{0.942$\pm$0.012} \\
 & MossFormer+Denoise (5 dB)$^\dagger$ & 0.325$\pm$0.039 & 0.989$\pm$0.004 & 0.995$\pm$0.000 & \textbf{0.996$\pm$0.000} & 0.987$\pm$0.004 \\
 & MossFormer+Denoise (10 dB)$^\dagger$ & 0.695$\pm$0.030 & 0.994$\pm$0.000 & 0.995$\pm$0.000 & \textbf{0.996$\pm$0.000} & 0.994$\pm$0.000 \\
 & MossFormer+Denoise (15 dB)$^\dagger$ & 0.899$\pm$0.016 & 0.994$\pm$0.000 & 0.995$\pm$0.000 & \textbf{0.996$\pm$0.000} & 0.994$\pm$0.000 \\
 & MossFormer+Denoise (20 dB)$^\dagger$ & 0.964$\pm$0.009 & 0.992$\pm$0.003 & 0.995$\pm$0.000 & \textbf{0.996$\pm$0.000} & 0.994$\pm$0.000 \\
\midrule
\multirow{2}{*}{\textbf{Vocoder}}
 & HiFi-GAN$^\dagger$ & 0.778$\pm$0.025 & 0.994$\pm$0.000 & \textbf{0.993$\pm$0.002} & \textbf{0.990$\pm$0.003} & \textbf{0.994$\pm$0.000} \\
 & Vocos$^\dagger$ & \textbf{0.998$\pm$0.000} & 0.994$\pm$0.000 & 0.995$\pm$0.000 & 0.996$\pm$0.000 & 0.994$\pm$0.000 \\
\bottomrule
\end{tabular}
\caption{Full ratio sweep of the component ablation under neural re-synthesis attacks (F1$\pm$SE at each incremental stage, 200-clip subset). $\dagger$ denotes unseen attacks. Supplementary version of the component ablation table above, part 2 of 2.}
\label{tab:supp_ablation_robustness_full_neural}
\end{table*}

\FloatBarrier

\begin{table*}[t]
\centering
\small
\setlength{\tabcolsep}{3pt}
\begin{tabular}{l l c c c c c c}
\toprule
\textbf{Category} & \textbf{Attack} & \textbf{WavMark} & \textbf{AudioSeal} & \textbf{TimbreWM} & \textbf{WMCodec} & \textbf{AWARE} & \textbf{\ours (Ours)} \\
\midrule
\multirow{57}{*}{\textbf{Sig.-Proc.}}
 & Crop Front (20\%) & 0.981$\pm$0.002 & 0.335$\pm$0.011 & 0.992$\pm$0.000 & 0.994$\pm$0.000 & \textbf{0.995$\pm$0.001} & \textbf{0.995$\pm$0.000} \\
 & Crop Front (40\%) & 0.973$\pm$0.002 & 0.324$\pm$0.011 & 0.992$\pm$0.000 & 0.986$\pm$0.001 & 0.969$\pm$0.002 & \textbf{0.995$\pm$0.000} \\
 & Crop Front (60\%) & 0.860$\pm$0.005 & 0.282$\pm$0.011 & 0.992$\pm$0.000 & 0.962$\pm$0.002 & 0.864$\pm$0.005 & \textbf{0.995$\pm$0.000} \\
 & Crop Front (80\%) & 0.562$\pm$0.010 & 0.371$\pm$0.011 & 0.992$\pm$0.000 & 0.840$\pm$0.006 & 0.491$\pm$0.011 & \textbf{0.994$\pm$0.000} \\
 & Crop Front (90\%) & 0.194$\pm$0.010 & 0.183$\pm$0.010 & 0.992$\pm$0.000 & 0.593$\pm$0.009 & 0.131$\pm$0.009 & \textbf{0.994$\pm$0.000} \\
 & Crop Front (95\%) & 0.014$\pm$0.003 & 0.101$\pm$0.008 & \textbf{0.992$\pm$0.000} & 0.308$\pm$0.011 & 0.022$\pm$0.004 & \textbf{0.992$\pm$0.001} \\
 & Crop Front (96\%) & 0.002$\pm$0.001 & 0.100$\pm$0.008 & \textbf{0.991$\pm$0.000} & 0.240$\pm$0.010 & 0.017$\pm$0.004 & \textbf{0.991$\pm$0.001} \\
 & Crop Front (97\%) & 0.000$\pm$0.000 & 0.058$\pm$0.006 & \textbf{0.986$\pm$0.001} & 0.160$\pm$0.009 & 0.012$\pm$0.003 & \textbf{0.988$\pm$0.001} \\
 & Crop Front (98\%) & 0.000$\pm$0.000 & 0.054$\pm$0.006 & 0.970$\pm$0.002 & 0.098$\pm$0.008 & 0.014$\pm$0.003 & \textbf{0.980$\pm$0.002} \\
 & Crop Middle (20\%) & \textbf{0.997$\pm$0.001} & 0.995$\pm$0.000 & 0.992$\pm$0.000 & 0.991$\pm$0.001 & 0.994$\pm$0.001 & 0.995$\pm$0.000 \\
 & Crop Middle (40\%) & 0.986$\pm$0.002 & \textbf{0.995$\pm$0.000} & 0.992$\pm$0.000 & 0.977$\pm$0.002 & 0.965$\pm$0.003 & 0.995$\pm$0.000 \\
 & Crop Middle (60\%) & 0.907$\pm$0.004 & \textbf{0.995$\pm$0.000} & 0.992$\pm$0.000 & 0.921$\pm$0.004 & 0.857$\pm$0.005 & 0.995$\pm$0.000 \\
 & Crop Middle (80\%) & 0.639$\pm$0.009 & \textbf{0.995$\pm$0.000} & 0.992$\pm$0.000 & 0.726$\pm$0.008 & 0.477$\pm$0.010 & 0.994$\pm$0.000 \\
 & Crop Middle (90\%) & 0.271$\pm$0.011 & \textbf{0.995$\pm$0.000} & 0.992$\pm$0.000 & 0.421$\pm$0.011 & 0.114$\pm$0.008 & 0.994$\pm$0.000 \\
 & Crop Middle (95\%) & 0.023$\pm$0.004 & \textbf{0.995$\pm$0.000} & 0.992$\pm$0.000 & 0.163$\pm$0.009 & 0.015$\pm$0.003 & 0.991$\pm$0.001 \\
 & Crop Middle (96\%) & 0.007$\pm$0.002 & \textbf{0.995$\pm$0.000} & 0.991$\pm$0.001 & 0.133$\pm$0.009 & 0.012$\pm$0.003 & 0.991$\pm$0.001 \\
 & Crop Middle (97\%) & 0.000$\pm$0.000 & \textbf{0.994$\pm$0.000} & 0.989$\pm$0.001 & 0.097$\pm$0.007 & 0.008$\pm$0.002 & 0.986$\pm$0.001 \\
 & Crop Middle (98\%) & 0.000$\pm$0.000 & \textbf{0.988$\pm$0.001} & 0.982$\pm$0.001 & 0.096$\pm$0.008 & 0.009$\pm$0.003 & 0.973$\pm$0.002 \\
 & Crop Back (20\%) & \textbf{1.000$\pm$0.000} & 0.995$\pm$0.000 & 0.992$\pm$0.000 & 0.994$\pm$0.000 & 0.997$\pm$0.000 & 0.995$\pm$0.000 \\
 & Crop Back (40\%) & \textbf{0.998$\pm$0.001} & 0.995$\pm$0.000 & 0.992$\pm$0.000 & 0.989$\pm$0.001 & 0.994$\pm$0.001 & 0.995$\pm$0.000 \\
 & Crop Back (60\%) & 0.954$\pm$0.003 & \textbf{0.995$\pm$0.000} & 0.992$\pm$0.000 & 0.966$\pm$0.002 & 0.967$\pm$0.002 & 0.995$\pm$0.000 \\
 & Crop Back (80\%) & 0.713$\pm$0.008 & \textbf{0.995$\pm$0.000} & 0.992$\pm$0.000 & 0.839$\pm$0.006 & 0.763$\pm$0.007 & 0.994$\pm$0.000 \\
 & Crop Back (90\%) & 0.353$\pm$0.011 & \textbf{0.995$\pm$0.000} & 0.992$\pm$0.000 & 0.593$\pm$0.010 & 0.380$\pm$0.010 & 0.991$\pm$0.001 \\
 & Crop Back (95\%) & 0.048$\pm$0.006 & \textbf{0.995$\pm$0.000} & 0.992$\pm$0.000 & 0.296$\pm$0.010 & 0.073$\pm$0.007 & 0.988$\pm$0.001 \\
 & Crop Back (96\%) & 0.015$\pm$0.003 & \textbf{0.995$\pm$0.000} & 0.992$\pm$0.000 & 0.212$\pm$0.010 & 0.041$\pm$0.005 & 0.985$\pm$0.001 \\
 & Crop Back (97\%) & 0.000$\pm$0.000 & \textbf{0.995$\pm$0.000} & 0.992$\pm$0.000 & 0.131$\pm$0.009 & 0.018$\pm$0.004 & 0.981$\pm$0.002 \\
 & Crop Back (98\%) & 0.000$\pm$0.000 & \textbf{0.995$\pm$0.000} & 0.992$\pm$0.000 & 0.072$\pm$0.007 & 0.010$\pm$0.003 & 0.970$\pm$0.002 \\
 & Resample 16kHz & \textbf{1.000$\pm$0.000} & 0.995$\pm$0.000 & 0.992$\pm$0.000 & 0.995$\pm$0.000 & 0.997$\pm$0.000 & 0.995$\pm$0.000 \\
 & Resample 8kHz & \textbf{1.000$\pm$0.000} & 0.995$\pm$0.000 & 0.992$\pm$0.000 & 0.738$\pm$0.008 & 0.997$\pm$0.000 & 0.995$\pm$0.000 \\
 & White Noise (20 dB) & 0.394$\pm$0.011 & 0.992$\pm$0.001 & 0.991$\pm$0.001 & 0.960$\pm$0.003 & \textbf{0.996$\pm$0.000} & 0.995$\pm$0.000 \\
 & White Noise (25 dB) & 0.833$\pm$0.006 & 0.995$\pm$0.000 & 0.992$\pm$0.000 & 0.990$\pm$0.001 & \textbf{0.997$\pm$0.000} & 0.995$\pm$0.000 \\
 & White Noise (30 dB) & 0.976$\pm$0.002 & 0.995$\pm$0.000 & 0.992$\pm$0.000 & 0.994$\pm$0.000 & \textbf{0.997$\pm$0.000} & 0.995$\pm$0.000 \\
 & White Noise (35 dB) & \textbf{0.996$\pm$0.001} & 0.995$\pm$0.000 & 0.992$\pm$0.000 & 0.994$\pm$0.000 & \textbf{0.997$\pm$0.000} & 0.995$\pm$0.000 \\
 & White Noise (40 dB) & \textbf{0.999$\pm$0.000} & 0.995$\pm$0.000 & 0.992$\pm$0.000 & 0.995$\pm$0.000 & 0.997$\pm$0.000 & 0.995$\pm$0.000 \\
 & Amp Scale (20\%) & \textbf{1.000$\pm$0.000} & 0.995$\pm$0.000 & 0.992$\pm$0.000 & 0.995$\pm$0.000 & 0.997$\pm$0.000 & 0.995$\pm$0.000 \\
 & Amp Scale (40\%) & \textbf{1.000$\pm$0.000} & 0.995$\pm$0.000 & 0.992$\pm$0.000 & 0.995$\pm$0.000 & 0.997$\pm$0.000 & 0.995$\pm$0.000 \\
 & Amp Scale (60\%) & \textbf{1.000$\pm$0.000} & 0.995$\pm$0.000 & 0.992$\pm$0.000 & 0.995$\pm$0.000 & 0.997$\pm$0.000 & 0.995$\pm$0.000 \\
 & Amp Scale (80\%) & \textbf{1.000$\pm$0.000} & 0.995$\pm$0.000 & 0.992$\pm$0.000 & 0.995$\pm$0.000 & 0.997$\pm$0.000 & 0.995$\pm$0.000 \\
 & MP3 Comp (8 kbps) & 0.000$\pm$0.000 & 0.858$\pm$0.005 & 0.885$\pm$0.005 & 0.039$\pm$0.005 & 0.458$\pm$0.011 & \textbf{0.994$\pm$0.000} \\
 & MP3 Comp (16 kbps) & 0.386$\pm$0.011 & \textbf{0.995$\pm$0.000} & 0.992$\pm$0.000 & 0.431$\pm$0.011 & \textbf{0.995$\pm$0.001} & 0.995$\pm$0.000 \\
 & MP3 Comp (24 kbps) & 0.962$\pm$0.003 & 0.995$\pm$0.000 & 0.992$\pm$0.000 & 0.884$\pm$0.005 & \textbf{0.997$\pm$0.000} & 0.995$\pm$0.000 \\
 & MP3 Comp (32 kbps) & \textbf{0.998$\pm$0.001} & 0.995$\pm$0.000 & 0.992$\pm$0.000 & 0.972$\pm$0.002 & \textbf{0.997$\pm$0.000} & 0.995$\pm$0.000 \\
 & MP3 Comp (40 kbps) & \textbf{1.000$\pm$0.000} & 0.995$\pm$0.000 & 0.992$\pm$0.000 & 0.989$\pm$0.001 & 0.997$\pm$0.000 & 0.995$\pm$0.000 \\
 & MP3 Comp (48 kbps) & \textbf{1.000$\pm$0.000} & 0.995$\pm$0.000 & 0.992$\pm$0.000 & 0.993$\pm$0.001 & 0.997$\pm$0.000 & 0.995$\pm$0.000 \\
 & MP3 Comp (56 kbps) & \textbf{1.000$\pm$0.000} & 0.995$\pm$0.000 & 0.992$\pm$0.000 & 0.994$\pm$0.000 & 0.997$\pm$0.000 & 0.995$\pm$0.000 \\
 & MP3 Comp (64 kbps) & \textbf{1.000$\pm$0.000} & 0.995$\pm$0.000 & 0.992$\pm$0.000 & 0.995$\pm$0.000 & 0.997$\pm$0.000 & 0.995$\pm$0.000 \\
 & Recount 8bps & 0.926$\pm$0.004 & 0.995$\pm$0.000 & 0.992$\pm$0.000 & 0.993$\pm$0.001 & \textbf{0.997$\pm$0.000} & 0.995$\pm$0.000 \\
 & Median Filt (5) & 0.993$\pm$0.001 & 0.995$\pm$0.000 & 0.992$\pm$0.000 & 0.866$\pm$0.005 & \textbf{0.997$\pm$0.000} & 0.995$\pm$0.000 \\
 & Median Filt (15) & 0.163$\pm$0.009 & \textbf{0.995$\pm$0.000} & 0.991$\pm$0.001 & 0.331$\pm$0.011 & 0.230$\pm$0.010 & 0.994$\pm$0.000 \\
 & Median Filt (25) & 0.055$\pm$0.006 & 0.361$\pm$0.011 & 0.978$\pm$0.002 & 0.120$\pm$0.008 & 0.047$\pm$0.006 & \textbf{0.991$\pm$0.001} \\
 & Median Filt (35) & 0.014$\pm$0.003 & 0.096$\pm$0.007 & 0.922$\pm$0.004 & 0.127$\pm$0.008 & 0.039$\pm$0.005 & \textbf{0.975$\pm$0.002} \\
 & Time Stretch (-20\%) & 0.950$\pm$0.003 & 0.982$\pm$0.002 & 0.992$\pm$0.000 & 0.888$\pm$0.005 & 0.970$\pm$0.002 & \textbf{0.995$\pm$0.000} \\
 & Time Stretch (-10\%) & 0.965$\pm$0.003 & 0.982$\pm$0.002 & 0.992$\pm$0.000 & 0.885$\pm$0.005 & 0.975$\pm$0.002 & \textbf{0.995$\pm$0.000} \\
 & Time Stretch (-5\%) & 0.975$\pm$0.002 & 0.979$\pm$0.002 & 0.992$\pm$0.000 & 0.882$\pm$0.005 & 0.982$\pm$0.002 & \textbf{0.995$\pm$0.000} \\
 & Time Stretch (5\%) & \textbf{0.993$\pm$0.001} & \textbf{0.995$\pm$0.000} & 0.992$\pm$0.000 & 0.905$\pm$0.004 & 0.985$\pm$0.001 & \textbf{0.995$\pm$0.000} \\
 & Time Stretch (10\%) & 0.973$\pm$0.002 & 0.993$\pm$0.001 & 0.992$\pm$0.000 & 0.867$\pm$0.005 & 0.970$\pm$0.002 & \textbf{0.995$\pm$0.000} \\
 & Time Stretch (20\%) & 0.907$\pm$0.004 & 0.989$\pm$0.001 & 0.992$\pm$0.000 & 0.843$\pm$0.006 & 0.954$\pm$0.003 & \textbf{0.994$\pm$0.000} \\
\midrule
\multirow{2}{*}{\textbf{Filter}}
 & LowPass 2kHz & 0.000$\pm$0.000 & \textbf{0.995$\pm$0.000} & 0.835$\pm$0.006 & 0.530$\pm$0.010 & 0.875$\pm$0.005 & 0.995$\pm$0.000 \\
 & HighPass 500Hz & \textbf{1.000$\pm$0.000} & 0.995$\pm$0.000 & 0.992$\pm$0.000 & 0.995$\pm$0.000 & 0.997$\pm$0.000 & 0.995$\pm$0.000 \\
\bottomrule
\end{tabular}
\caption{Full ratio sweep of the main robustness comparison under traditional signal-level distortions (F1$\pm$SE across attacks, all models). Supplementary version of the baseline comparison table above, part 1 of 2.}
\label{tab:supp_main_robustness_full_sigproc}
\end{table*}

\FloatBarrier

\begin{table*}[t]
\centering
\small
\setlength{\tabcolsep}{3pt}
\begin{tabular}{l l c c c c c c}
\toprule
\textbf{Category} & \textbf{Attack} & \textbf{WavMark} & \textbf{AudioSeal} & \textbf{TimbreWM} & \textbf{WMCodec} & \textbf{AWARE} & \textbf{\ours (Ours)} \\
\midrule
\multirow{3}{*}{\textbf{Codec}}
 & FACodec & 0.000$\pm$0.000 & 0.021$\pm$0.004 & 0.041$\pm$0.005 & 0.019$\pm$0.004 & 0.149$\pm$0.009 & \textbf{0.974$\pm$0.002} \\
 & EnCodec 6kbps$^\dagger$ & 0.000$\pm$0.000 & 0.150$\pm$0.009 & 0.072$\pm$0.007 & 0.033$\pm$0.005 & 0.252$\pm$0.010 & \textbf{0.987$\pm$0.001} \\
 & TiCodec 1g4r$^\dagger$ & 0.000$\pm$0.000 & 0.023$\pm$0.004 & 0.033$\pm$0.005 & 0.027$\pm$0.004 & 0.075$\pm$0.007 & \textbf{0.826$\pm$0.006} \\
\midrule
\multirow{10}{*}{\textbf{Denoiser}}
 & FRCRN+Denoise (0 dB)$^\dagger$ & 0.000$\pm$0.000 & 0.034$\pm$0.005 & 0.117$\pm$0.008 & 0.073$\pm$0.007 & 0.629$\pm$0.009 & \textbf{0.944$\pm$0.003} \\
 & FRCRN+Denoise (5 dB)$^\dagger$ & 0.000$\pm$0.000 & 0.079$\pm$0.007 & 0.437$\pm$0.011 & 0.224$\pm$0.010 & 0.863$\pm$0.005 & \textbf{0.991$\pm$0.001} \\
 & FRCRN+Denoise (10 dB)$^\dagger$ & 0.000$\pm$0.000 & 0.342$\pm$0.011 & 0.814$\pm$0.006 & 0.536$\pm$0.010 & 0.967$\pm$0.002 & \textbf{0.994$\pm$0.000} \\
 & FRCRN+Denoise (15 dB)$^\dagger$ & 0.000$\pm$0.000 & 0.739$\pm$0.008 & 0.962$\pm$0.002 & 0.833$\pm$0.006 & 0.992$\pm$0.001 & \textbf{0.994$\pm$0.000} \\
 & FRCRN+Denoise (20 dB)$^\dagger$ & 0.047$\pm$0.006 & 0.940$\pm$0.003 & 0.990$\pm$0.001 & 0.956$\pm$0.003 & \textbf{0.996$\pm$0.000} & 0.995$\pm$0.000 \\
 & MossFormer+Denoise (0 dB)$^\dagger$ & 0.000$\pm$0.000 & 0.029$\pm$0.004 & 0.107$\pm$0.008 & 0.111$\pm$0.008 & 0.614$\pm$0.009 & \textbf{0.956$\pm$0.003} \\
 & MossFormer+Denoise (5 dB)$^\dagger$ & 0.000$\pm$0.000 & 0.073$\pm$0.007 & 0.368$\pm$0.011 & 0.281$\pm$0.011 & 0.860$\pm$0.005 & \textbf{0.991$\pm$0.001} \\
 & MossFormer+Denoise (10 dB)$^\dagger$ & 0.000$\pm$0.000 & 0.296$\pm$0.011 & 0.717$\pm$0.008 & 0.554$\pm$0.010 & 0.965$\pm$0.002 & \textbf{0.994$\pm$0.000} \\
 & MossFormer+Denoise (15 dB)$^\dagger$ & 0.001$\pm$0.001 & 0.670$\pm$0.009 & 0.910$\pm$0.004 & 0.813$\pm$0.006 & 0.991$\pm$0.001 & \textbf{0.994$\pm$0.000} \\
 & MossFormer+Denoise (20 dB)$^\dagger$ & 0.036$\pm$0.005 & 0.910$\pm$0.004 & 0.975$\pm$0.002 & 0.950$\pm$0.003 & \textbf{0.996$\pm$0.000} & 0.995$\pm$0.000 \\
\midrule
\multirow{2}{*}{\textbf{Vocoder}}
 & HiFi-GAN$^\dagger$ & 0.000$\pm$0.000 & 0.014$\pm$0.003 & 0.809$\pm$0.006 & 0.302$\pm$0.010 & 0.713$\pm$0.008 & \textbf{0.995$\pm$0.000} \\
 & Vocos$^\dagger$ & 0.000$\pm$0.000 & 0.029$\pm$0.005 & 0.992$\pm$0.000 & 0.994$\pm$0.000 & \textbf{0.996$\pm$0.000} & 0.994$\pm$0.000 \\
\bottomrule
\end{tabular}
\caption{Full ratio sweep of the main robustness comparison under neural re-synthesis attacks (F1$\pm$SE across attacks, all models). $\dagger$ denotes unseen attacks. Supplementary version of the baseline comparison table above, part 2 of 2.}
\label{tab:supp_main_robustness_full_neural}
\end{table*}

\FloatBarrier

\begin{table*}[t]
\centering
\small
\setlength{\tabcolsep}{4pt}
\begin{tabular}{l l c c c c c}
\toprule
\textbf{Category} & \textbf{Attack} & \textbf{No ECC} & \textbf{Rep3} & \textbf{RS(4,2)} & \textbf{RS(6,2)} & \textbf{LDPC} \\
\midrule
\multirow{57}{*}{\textbf{Sig.-Proc.}}
 & Crop Front (20\%) & \textbf{0.996$\pm$0.000} & 0.994$\pm$0.000 & 0.995$\pm$0.000 & 0.994$\pm$0.000 & 0.995$\pm$0.000 \\
 & Crop Front (40\%) & \textbf{0.993$\pm$0.002} & 0.994$\pm$0.000 & 0.995$\pm$0.000 & 0.994$\pm$0.000 & \textbf{0.995$\pm$0.000} \\
 & Crop Front (60\%) & \textbf{0.993$\pm$0.002} & 0.994$\pm$0.000 & 0.995$\pm$0.000 & 0.994$\pm$0.000 & \textbf{0.995$\pm$0.000} \\
 & Crop Front (80\%) & 0.980$\pm$0.006 & 0.994$\pm$0.000 & 0.995$\pm$0.000 & 0.994$\pm$0.000 & \textbf{0.995$\pm$0.000} \\
 & Crop Front (90\%) & 0.959$\pm$0.010 & 0.994$\pm$0.000 & \textbf{0.995$\pm$0.000} & 0.994$\pm$0.000 & \textbf{0.993$\pm$0.003} \\
 & Crop Front (95\%) & 0.965$\pm$0.009 & \textbf{0.994$\pm$0.000} & \textbf{0.990$\pm$0.003} & 0.992$\pm$0.002 & 0.972$\pm$0.007 \\
 & Crop Front (96\%) & 0.962$\pm$0.009 & \textbf{0.994$\pm$0.000} & \textbf{0.992$\pm$0.003} & 0.994$\pm$0.000 & 0.982$\pm$0.006 \\
 & Crop Front (97\%) & 0.983$\pm$0.006 & \textbf{0.994$\pm$0.000} & \textbf{0.992$\pm$0.003} & 0.992$\pm$0.003 & 0.980$\pm$0.006 \\
 & Crop Front (98\%) & \textbf{0.985$\pm$0.005} & \textbf{0.984$\pm$0.005} & 0.959$\pm$0.010 & \textbf{0.979$\pm$0.006} & 0.942$\pm$0.012 \\
 & Crop Middle (20\%) & \textbf{0.996$\pm$0.000} & 0.994$\pm$0.000 & 0.995$\pm$0.000 & 0.994$\pm$0.000 & 0.995$\pm$0.000 \\
 & Crop Middle (40\%) & \textbf{0.996$\pm$0.000} & 0.994$\pm$0.000 & 0.995$\pm$0.000 & 0.994$\pm$0.000 & 0.995$\pm$0.000 \\
 & Crop Middle (60\%) & \textbf{0.996$\pm$0.000} & 0.994$\pm$0.000 & 0.995$\pm$0.000 & 0.994$\pm$0.000 & 0.995$\pm$0.000 \\
 & Crop Middle (80\%) & \textbf{0.996$\pm$0.000} & 0.994$\pm$0.000 & 0.995$\pm$0.000 & 0.994$\pm$0.000 & 0.995$\pm$0.000 \\
 & Crop Middle (90\%) & \textbf{0.996$\pm$0.000} & 0.994$\pm$0.000 & 0.995$\pm$0.000 & 0.994$\pm$0.000 & 0.995$\pm$0.000 \\
 & Crop Middle (95\%) & \textbf{0.996$\pm$0.000} & 0.992$\pm$0.003 & 0.995$\pm$0.000 & 0.994$\pm$0.000 & 0.995$\pm$0.000 \\
 & Crop Middle (96\%) & \textbf{0.996$\pm$0.000} & 0.994$\pm$0.000 & 0.992$\pm$0.003 & 0.992$\pm$0.003 & 0.993$\pm$0.003 \\
 & Crop Middle (97\%) & \textbf{0.996$\pm$0.000} & 0.987$\pm$0.004 & 0.992$\pm$0.003 & 0.992$\pm$0.003 & 0.972$\pm$0.008 \\
 & Crop Middle (98\%) & \textbf{0.993$\pm$0.002} & 0.974$\pm$0.007 & 0.974$\pm$0.007 & 0.966$\pm$0.008 & 0.928$\pm$0.014 \\
 & Crop Back (20\%) & \textbf{0.996$\pm$0.000} & 0.994$\pm$0.000 & 0.995$\pm$0.000 & 0.994$\pm$0.000 & 0.995$\pm$0.000 \\
 & Crop Back (40\%) & \textbf{0.996$\pm$0.000} & 0.994$\pm$0.000 & 0.995$\pm$0.000 & 0.994$\pm$0.000 & 0.995$\pm$0.000 \\
 & Crop Back (60\%) & \textbf{0.996$\pm$0.000} & 0.994$\pm$0.000 & 0.995$\pm$0.000 & 0.994$\pm$0.000 & 0.995$\pm$0.000 \\
 & Crop Back (80\%) & \textbf{0.996$\pm$0.000} & 0.994$\pm$0.000 & 0.995$\pm$0.000 & 0.994$\pm$0.000 & 0.995$\pm$0.000 \\
 & Crop Back (90\%) & \textbf{0.993$\pm$0.003} & 0.992$\pm$0.003 & 0.992$\pm$0.003 & 0.992$\pm$0.003 & \textbf{0.995$\pm$0.000} \\
 & Crop Back (95\%) & \textbf{0.993$\pm$0.003} & 0.992$\pm$0.003 & 0.992$\pm$0.003 & 0.992$\pm$0.003 & \textbf{0.995$\pm$0.000} \\
 & Crop Back (96\%) & \textbf{0.990$\pm$0.003} & 0.984$\pm$0.005 & \textbf{0.992$\pm$0.003} & 0.989$\pm$0.003 & \textbf{0.990$\pm$0.004} \\
 & Crop Back (97\%) & \textbf{0.988$\pm$0.004} & 0.974$\pm$0.007 & \textbf{0.987$\pm$0.004} & \textbf{0.986$\pm$0.004} & 0.977$\pm$0.007 \\
 & Crop Back (98\%) & \textbf{0.985$\pm$0.005} & 0.969$\pm$0.008 & \textbf{0.982$\pm$0.005} & 0.979$\pm$0.006 & 0.937$\pm$0.013 \\
 & Resample 16kHz & \textbf{0.996$\pm$0.000} & 0.994$\pm$0.000 & 0.995$\pm$0.000 & 0.994$\pm$0.000 & 0.995$\pm$0.000 \\
 & Resample 8kHz & \textbf{0.996$\pm$0.000} & 0.994$\pm$0.000 & 0.995$\pm$0.000 & 0.994$\pm$0.000 & 0.995$\pm$0.000 \\
 & White Noise (20 dB) & \textbf{0.996$\pm$0.000} & 0.994$\pm$0.000 & 0.995$\pm$0.000 & 0.994$\pm$0.000 & 0.995$\pm$0.000 \\
 & White Noise (25 dB) & \textbf{0.996$\pm$0.000} & 0.994$\pm$0.000 & 0.995$\pm$0.000 & 0.994$\pm$0.000 & 0.995$\pm$0.000 \\
 & White Noise (30 dB) & \textbf{0.996$\pm$0.000} & 0.994$\pm$0.000 & 0.995$\pm$0.000 & 0.994$\pm$0.000 & 0.995$\pm$0.000 \\
 & White Noise (35 dB) & \textbf{0.996$\pm$0.000} & 0.994$\pm$0.000 & 0.995$\pm$0.000 & 0.994$\pm$0.000 & 0.995$\pm$0.000 \\
 & White Noise (40 dB) & \textbf{0.996$\pm$0.000} & 0.994$\pm$0.000 & 0.995$\pm$0.000 & 0.994$\pm$0.000 & 0.995$\pm$0.000 \\
 & Amp Scale (20\%) & \textbf{0.996$\pm$0.000} & 0.994$\pm$0.000 & 0.995$\pm$0.000 & 0.994$\pm$0.000 & 0.995$\pm$0.000 \\
 & Amp Scale (40\%) & \textbf{0.996$\pm$0.000} & 0.994$\pm$0.000 & 0.995$\pm$0.000 & 0.994$\pm$0.000 & 0.995$\pm$0.000 \\
 & Amp Scale (60\%) & \textbf{0.996$\pm$0.000} & 0.994$\pm$0.000 & 0.995$\pm$0.000 & 0.994$\pm$0.000 & 0.995$\pm$0.000 \\
 & Amp Scale (80\%) & \textbf{0.996$\pm$0.000} & 0.994$\pm$0.000 & 0.995$\pm$0.000 & 0.994$\pm$0.000 & 0.995$\pm$0.000 \\
 & MP3 Comp (8 kbps) & \textbf{0.996$\pm$0.000} & 0.994$\pm$0.000 & 0.995$\pm$0.000 & 0.992$\pm$0.002 & 0.995$\pm$0.000 \\
 & MP3 Comp (16 kbps) & \textbf{0.996$\pm$0.000} & 0.994$\pm$0.000 & 0.995$\pm$0.000 & 0.994$\pm$0.000 & 0.995$\pm$0.000 \\
 & MP3 Comp (24 kbps) & \textbf{0.996$\pm$0.000} & 0.994$\pm$0.000 & 0.995$\pm$0.000 & 0.994$\pm$0.000 & 0.995$\pm$0.000 \\
 & MP3 Comp (32 kbps) & \textbf{0.996$\pm$0.000} & 0.994$\pm$0.000 & 0.995$\pm$0.000 & 0.994$\pm$0.000 & 0.995$\pm$0.000 \\
 & MP3 Comp (40 kbps) & \textbf{0.996$\pm$0.000} & 0.994$\pm$0.000 & 0.995$\pm$0.000 & 0.994$\pm$0.000 & 0.995$\pm$0.000 \\
 & MP3 Comp (48 kbps) & \textbf{0.996$\pm$0.000} & 0.994$\pm$0.000 & 0.995$\pm$0.000 & 0.994$\pm$0.000 & 0.995$\pm$0.000 \\
 & MP3 Comp (56 kbps) & \textbf{0.996$\pm$0.000} & 0.994$\pm$0.000 & 0.995$\pm$0.000 & 0.994$\pm$0.000 & 0.995$\pm$0.000 \\
 & MP3 Comp (64 kbps) & \textbf{0.996$\pm$0.000} & 0.994$\pm$0.000 & 0.995$\pm$0.000 & 0.994$\pm$0.000 & 0.995$\pm$0.000 \\
 & Recount 8bps & \textbf{0.996$\pm$0.000} & 0.994$\pm$0.000 & 0.995$\pm$0.000 & 0.994$\pm$0.000 & 0.995$\pm$0.000 \\
 & Median Filt (5) & \textbf{0.996$\pm$0.000} & 0.994$\pm$0.000 & 0.995$\pm$0.000 & 0.994$\pm$0.000 & 0.995$\pm$0.000 \\
 & Median Filt (15) & \textbf{0.996$\pm$0.000} & 0.994$\pm$0.000 & 0.995$\pm$0.000 & 0.992$\pm$0.003 & 0.995$\pm$0.000 \\
 & Median Filt (25) & \textbf{0.996$\pm$0.000} & 0.989$\pm$0.003 & 0.980$\pm$0.006 & 0.958$\pm$0.010 & 0.969$\pm$0.008 \\
 & Median Filt (35) & \textbf{0.983$\pm$0.005} & \textbf{0.976$\pm$0.007} & 0.928$\pm$0.014 & 0.925$\pm$0.014 & 0.920$\pm$0.015 \\
 & Time Stretch (-20\%) & \textbf{0.996$\pm$0.000} & 0.994$\pm$0.000 & 0.995$\pm$0.000 & 0.994$\pm$0.000 & 0.995$\pm$0.000 \\
 & Time Stretch (-10\%) & \textbf{0.996$\pm$0.000} & 0.994$\pm$0.000 & 0.995$\pm$0.000 & 0.994$\pm$0.000 & 0.995$\pm$0.000 \\
 & Time Stretch (-5\%) & \textbf{0.996$\pm$0.000} & 0.994$\pm$0.000 & 0.995$\pm$0.000 & 0.994$\pm$0.000 & 0.995$\pm$0.000 \\
 & Time Stretch (5\%) & \textbf{0.996$\pm$0.000} & 0.994$\pm$0.000 & 0.995$\pm$0.000 & 0.994$\pm$0.000 & 0.995$\pm$0.000 \\
 & Time Stretch (10\%) & \textbf{0.996$\pm$0.000} & 0.994$\pm$0.000 & 0.995$\pm$0.000 & 0.994$\pm$0.000 & 0.995$\pm$0.000 \\
 & Time Stretch (20\%) & \textbf{0.996$\pm$0.000} & 0.994$\pm$0.000 & 0.995$\pm$0.000 & 0.994$\pm$0.000 & 0.995$\pm$0.000 \\
\midrule
\multirow{2}{*}{\textbf{Filter}}
 & LowPass 2kHz & \textbf{0.996$\pm$0.000} & 0.994$\pm$0.000 & 0.995$\pm$0.000 & 0.994$\pm$0.000 & 0.993$\pm$0.002 \\
 & HighPass 500Hz & \textbf{0.996$\pm$0.000} & 0.994$\pm$0.000 & 0.995$\pm$0.000 & 0.994$\pm$0.000 & 0.995$\pm$0.000 \\
\bottomrule
\end{tabular}
\caption{Full ratio sweep of ECC variant robustness under traditional signal-level distortions (F1$\pm$SE, 200-clip subset), referenced in the Error-Correcting Code paragraph of Ablation Studies above, part 1 of 2.}
\label{tab:supp_ecc_robustness_full_sigproc}
\end{table*}

\FloatBarrier

\begin{table*}[t]
\centering
\small
\setlength{\tabcolsep}{4pt}
\begin{tabular}{l l c c c c c}
\toprule
\textbf{Category} & \textbf{Attack} & \textbf{No ECC} & \textbf{Rep3} & \textbf{RS(4,2)} & \textbf{RS(6,2)} & \textbf{LDPC} \\
\midrule
\multirow{3}{*}{\textbf{Codec}}
 & FACodec & 0.937$\pm$0.012 & \textbf{0.982$\pm$0.005} & \textbf{0.969$\pm$0.008} & 0.835$\pm$0.020 & 0.945$\pm$0.012 \\
 & EnCodec 6kbps$^\dagger$ & 0.926$\pm$0.014 & \textbf{0.982$\pm$0.006} & 0.902$\pm$0.016 & 0.486$\pm$0.038 & 0.839$\pm$0.021 \\
 & TiCodec 1g4r$^\dagger$ & 0.684$\pm$0.030 & \textbf{0.838$\pm$0.021} & 0.611$\pm$0.033 & 0.288$\pm$0.038 & 0.428$\pm$0.039 \\
\midrule
\multirow{10}{*}{\textbf{Denoiser}}
 & FRCRN+Denoise (0 dB)$^\dagger$ & 0.943$\pm$0.012 & 0.913$\pm$0.015 & \textbf{0.977$\pm$0.007} & \textbf{0.976$\pm$0.007} & 0.931$\pm$0.014 \\
 & FRCRN+Denoise (5 dB)$^\dagger$ & \textbf{0.988$\pm$0.004} & 0.989$\pm$0.003 & \textbf{0.992$\pm$0.003} & 0.992$\pm$0.003 & \textbf{0.985$\pm$0.005} \\
 & FRCRN+Denoise (10 dB)$^\dagger$ & \textbf{0.996$\pm$0.000} & 0.992$\pm$0.003 & 0.995$\pm$0.000 & 0.994$\pm$0.000 & 0.993$\pm$0.002 \\
 & FRCRN+Denoise (15 dB)$^\dagger$ & \textbf{0.996$\pm$0.000} & 0.994$\pm$0.000 & 0.995$\pm$0.000 & 0.994$\pm$0.000 & 0.995$\pm$0.000 \\
 & FRCRN+Denoise (20 dB)$^\dagger$ & \textbf{0.996$\pm$0.000} & 0.994$\pm$0.000 & 0.995$\pm$0.000 & 0.994$\pm$0.000 & 0.995$\pm$0.000 \\
 & MossFormer+Denoise (0 dB)$^\dagger$ & 0.954$\pm$0.010 & 0.942$\pm$0.012 & \textbf{0.985$\pm$0.005} & \textbf{0.981$\pm$0.006} & 0.951$\pm$0.011 \\
 & MossFormer+Denoise (5 dB)$^\dagger$ & \textbf{0.996$\pm$0.000} & 0.987$\pm$0.004 & 0.992$\pm$0.002 & 0.994$\pm$0.000 & 0.993$\pm$0.002 \\
 & MossFormer+Denoise (10 dB)$^\dagger$ & \textbf{0.996$\pm$0.000} & 0.994$\pm$0.000 & 0.995$\pm$0.000 & 0.994$\pm$0.000 & 0.995$\pm$0.000 \\
 & MossFormer+Denoise (15 dB)$^\dagger$ & \textbf{0.996$\pm$0.000} & 0.994$\pm$0.000 & 0.995$\pm$0.000 & 0.994$\pm$0.000 & 0.995$\pm$0.000 \\
 & MossFormer+Denoise (20 dB)$^\dagger$ & \textbf{0.996$\pm$0.000} & 0.994$\pm$0.000 & 0.995$\pm$0.000 & 0.994$\pm$0.000 & 0.995$\pm$0.000 \\
\midrule
\multirow{2}{*}{\textbf{Vocoder}}
 & HiFi-GAN$^\dagger$ & \textbf{0.990$\pm$0.003} & \textbf{0.994$\pm$0.000} & \textbf{0.990$\pm$0.004} & 0.855$\pm$0.020 & 0.977$\pm$0.007 \\
 & Vocos$^\dagger$ & \textbf{0.996$\pm$0.000} & 0.994$\pm$0.000 & 0.995$\pm$0.000 & 0.994$\pm$0.000 & 0.995$\pm$0.000 \\
\bottomrule
\end{tabular}
\caption{Full ratio sweep of ECC variant robustness under neural re-synthesis attacks (F1$\pm$SE, 200-clip subset). $\dagger$ denotes unseen attacks. Referenced in the Error-Correcting Code paragraph of Ablation Studies above, part 2 of 2.}
\label{tab:supp_ecc_robustness_full_neural}
\end{table*}

\FloatBarrier

\begin{table*}[t]
\centering
\small
\setlength{\tabcolsep}{4pt}
\begin{tabular}{l l c c}
\toprule
\textbf{Category} & \textbf{Attack} & \textbf{\ours} & \textbf{Zero-shot (LJSpeech)} \\
\midrule
\multirow{57}{*}{\textbf{Sig.-Proc.}}
 & Crop Front (20\%) & 0.994$\pm$0.000 & \textbf{0.995$\pm$0.000} \\
 & Crop Front (40\%) & 0.994$\pm$0.000 & \textbf{0.995$\pm$0.000} \\
 & Crop Front (60\%) & 0.994$\pm$0.000 & \textbf{0.995$\pm$0.000} \\
 & Crop Front (80\%) & 0.994$\pm$0.000 & \textbf{0.995$\pm$0.000} \\
 & Crop Front (90\%) & 0.994$\pm$0.000 & \textbf{0.995$\pm$0.000} \\
 & Crop Front (95\%) & 0.994$\pm$0.000 & \textbf{0.995$\pm$0.000} \\
 & Crop Front (96\%) & 0.994$\pm$0.000 & \textbf{0.995$\pm$0.000} \\
 & Crop Front (97\%) & 0.994$\pm$0.000 & \textbf{0.995$\pm$0.000} \\
 & Crop Front (98\%) & \textbf{0.984$\pm$0.005} & \textbf{0.984$\pm$0.005} \\
 & Crop Middle (20\%) & 0.994$\pm$0.000 & \textbf{0.995$\pm$0.000} \\
 & Crop Middle (40\%) & 0.994$\pm$0.000 & \textbf{0.995$\pm$0.000} \\
 & Crop Middle (60\%) & 0.994$\pm$0.000 & \textbf{0.995$\pm$0.000} \\
 & Crop Middle (80\%) & 0.994$\pm$0.000 & \textbf{0.995$\pm$0.000} \\
 & Crop Middle (90\%) & 0.994$\pm$0.000 & \textbf{0.995$\pm$0.000} \\
 & Crop Middle (95\%) & \textbf{0.992$\pm$0.003} & 0.950$\pm$0.010 \\
 & Crop Middle (96\%) & \textbf{0.994$\pm$0.000} & 0.928$\pm$0.013 \\
 & Crop Middle (97\%) & \textbf{0.987$\pm$0.004} & 0.871$\pm$0.019 \\
 & Crop Middle (98\%) & \textbf{0.974$\pm$0.007} & 0.671$\pm$0.031 \\
 & Crop Back (20\%) & 0.994$\pm$0.000 & \textbf{0.995$\pm$0.000} \\
 & Crop Back (40\%) & 0.994$\pm$0.000 & \textbf{0.995$\pm$0.000} \\
 & Crop Back (60\%) & 0.994$\pm$0.000 & \textbf{0.995$\pm$0.000} \\
 & Crop Back (80\%) & 0.994$\pm$0.000 & \textbf{0.995$\pm$0.000} \\
 & Crop Back (90\%) & \textbf{0.992$\pm$0.003} & \textbf{0.992$\pm$0.002} \\
 & Crop Back (95\%) & \textbf{0.992$\pm$0.003} & \textbf{0.992$\pm$0.003} \\
 & Crop Back (96\%) & \textbf{0.984$\pm$0.005} & \textbf{0.984$\pm$0.005} \\
 & Crop Back (97\%) & \textbf{0.974$\pm$0.007} & \textbf{0.958$\pm$0.010} \\
 & Crop Back (98\%) & \textbf{0.969$\pm$0.008} & 0.890$\pm$0.017 \\
 & Resample 16kHz & 0.994$\pm$0.000 & \textbf{0.995$\pm$0.000} \\
 & Resample 8kHz & 0.994$\pm$0.000 & \textbf{0.995$\pm$0.000} \\
 & White Noise (20 dB) & 0.994$\pm$0.000 & \textbf{0.995$\pm$0.000} \\
 & White Noise (25 dB) & 0.994$\pm$0.000 & \textbf{0.995$\pm$0.000} \\
 & White Noise (30 dB) & 0.994$\pm$0.000 & \textbf{0.995$\pm$0.000} \\
 & White Noise (35 dB) & 0.994$\pm$0.000 & \textbf{0.995$\pm$0.000} \\
 & White Noise (40 dB) & 0.994$\pm$0.000 & \textbf{0.995$\pm$0.000} \\
 & Amp Scale (20\%) & 0.994$\pm$0.000 & \textbf{0.995$\pm$0.000} \\
 & Amp Scale (40\%) & 0.994$\pm$0.000 & \textbf{0.995$\pm$0.000} \\
 & Amp Scale (60\%) & 0.994$\pm$0.000 & \textbf{0.995$\pm$0.000} \\
 & Amp Scale (80\%) & 0.994$\pm$0.000 & \textbf{0.995$\pm$0.000} \\
 & MP3 Comp (8 kbps) & 0.994$\pm$0.000 & \textbf{0.995$\pm$0.000} \\
 & MP3 Comp (16 kbps) & 0.994$\pm$0.000 & \textbf{0.995$\pm$0.000} \\
 & MP3 Comp (24 kbps) & 0.994$\pm$0.000 & \textbf{0.995$\pm$0.000} \\
 & MP3 Comp (32 kbps) & 0.994$\pm$0.000 & \textbf{0.995$\pm$0.000} \\
 & MP3 Comp (40 kbps) & 0.994$\pm$0.000 & \textbf{0.995$\pm$0.000} \\
 & MP3 Comp (48 kbps) & 0.994$\pm$0.000 & \textbf{0.995$\pm$0.000} \\
 & MP3 Comp (56 kbps) & 0.994$\pm$0.000 & \textbf{0.995$\pm$0.000} \\
 & MP3 Comp (64 kbps) & 0.994$\pm$0.000 & \textbf{0.995$\pm$0.000} \\
 & Recount 8bps & 0.994$\pm$0.000 & \textbf{0.995$\pm$0.000} \\
 & Median Filt (5) & 0.994$\pm$0.000 & \textbf{0.995$\pm$0.000} \\
 & Median Filt (15) & 0.994$\pm$0.000 & \textbf{0.995$\pm$0.000} \\
 & Median Filt (25) & 0.989$\pm$0.003 & \textbf{0.995$\pm$0.000} \\
 & Median Filt (35) & 0.976$\pm$0.007 & \textbf{0.990$\pm$0.004} \\
 & Time Stretch (-20\%) & 0.994$\pm$0.000 & \textbf{0.995$\pm$0.000} \\
 & Time Stretch (-10\%) & 0.994$\pm$0.000 & \textbf{0.995$\pm$0.000} \\
 & Time Stretch (-5\%) & 0.994$\pm$0.000 & \textbf{0.995$\pm$0.000} \\
 & Time Stretch (5\%) & 0.994$\pm$0.000 & \textbf{0.995$\pm$0.000} \\
 & Time Stretch (10\%) & 0.994$\pm$0.000 & \textbf{0.995$\pm$0.000} \\
 & Time Stretch (20\%) & 0.994$\pm$0.000 & \textbf{0.995$\pm$0.000} \\
\midrule
\multirow{2}{*}{\textbf{Filter}}
 & LowPass 2kHz & 0.994$\pm$0.000 & \textbf{0.995$\pm$0.000} \\
 & HighPass 500Hz & 0.994$\pm$0.000 & \textbf{0.995$\pm$0.000} \\
\bottomrule
\end{tabular}
\caption{Full ratio sweep of zero-shot generalization to LJSpeech under traditional signal-level distortions (F1$\pm$SE, 200-clip subset per domain), referenced in the Cross-Domain Generalization subsection above, part 1 of 2.}
\label{tab:supp_zeroshot_full_sigproc}
\end{table*}

\FloatBarrier

\begin{table*}[t]
\centering
\small
\setlength{\tabcolsep}{4pt}
\begin{tabular}{l l c c}
\toprule
\textbf{Category} & \textbf{Attack} & \textbf{\ours} & \textbf{Zero-shot (LJSpeech)} \\
\midrule
\multirow{3}{*}{\textbf{Codec}}
 & FACodec & 0.982$\pm$0.005 & \textbf{0.992$\pm$0.002} \\
 & EnCodec 6kbps & 0.982$\pm$0.006 & \textbf{0.995$\pm$0.000} \\
 & TiCodec 1g4r & \textbf{0.838$\pm$0.021} & \textbf{0.794$\pm$0.025} \\
\midrule
\multirow{10}{*}{\textbf{Denoiser}}
 & FRCRN+Denoise (0 dB) & 0.913$\pm$0.015 & \textbf{0.966$\pm$0.008} \\
 & FRCRN+Denoise (5 dB) & 0.989$\pm$0.003 & \textbf{0.995$\pm$0.000} \\
 & FRCRN+Denoise (10 dB) & 0.992$\pm$0.003 & \textbf{0.995$\pm$0.000} \\
 & FRCRN+Denoise (15 dB) & 0.994$\pm$0.000 & \textbf{0.995$\pm$0.000} \\
 & FRCRN+Denoise (20 dB) & 0.994$\pm$0.000 & \textbf{0.995$\pm$0.000} \\
 & MossFormer+Denoise (0 dB) & 0.942$\pm$0.012 & \textbf{0.974$\pm$0.007} \\
 & MossFormer+Denoise (5 dB) & 0.987$\pm$0.004 & \textbf{0.995$\pm$0.000} \\
 & MossFormer+Denoise (10 dB) & 0.994$\pm$0.000 & \textbf{0.995$\pm$0.000} \\
 & MossFormer+Denoise (15 dB) & 0.994$\pm$0.000 & \textbf{0.995$\pm$0.000} \\
 & MossFormer+Denoise (20 dB) & 0.994$\pm$0.000 & \textbf{0.995$\pm$0.000} \\
\midrule
\multirow{2}{*}{\textbf{Vocoder}}
 & HiFi-GAN & 0.994$\pm$0.000 & \textbf{0.995$\pm$0.000} \\
 & Vocos & 0.994$\pm$0.000 & \textbf{0.995$\pm$0.000} \\
\bottomrule
\end{tabular}
\caption{Full ratio sweep of zero-shot generalization to LJSpeech under neural re-synthesis attacks (F1$\pm$SE, 200-clip subset per domain), referenced in the Cross-Domain Generalization subsection above, part 2 of 2.}
\label{tab:supp_zeroshot_full_neural}
\end{table*}

\FloatBarrier

\end{document}